\documentclass[12pt,a4paper]{article}
\pdfoutput=1

\usepackage{geometry}
\usepackage{amsmath}
\usepackage{amssymb}
\usepackage[dvips]{graphicx}
\usepackage{cite}
\usepackage{pstricks}
\usepackage{bm}
\usepackage{pbox}
\usepackage{placeins}
\usepackage{graphicx}
\usepackage{caption}
\usepackage{subcaption}
\usepackage[T1]{fontenc}
\usepackage{footnote}
\usepackage{pdfpages}
\usepackage{hhline}
\usepackage{multirow}
\usepackage{multicol}
\usepackage{enumitem}
\usepackage[toc,page]{appendix}
\allowdisplaybreaks

\definecolor{MyDarkBlue}{rgb}{0.1, 0.1, 0.8} 
\definecolor{MyLightBlue}{rgb}{0.22,0.51,0.9}
\definecolor{MyGreen}{rgb}{0.0, 0.45, 0.0}
\definecolor{BrickRed}{rgb}{0.8, 0.25, 0.33}
\usepackage[colorlinks=true,linkcolor=blue,citecolor=MyDarkBlue,
urlcolor=MyLightBlue,bookmarksnumbered=true,bookmarksopen]{hyperref}
\hypersetup{colorlinks, citecolor=BrickRed,linkcolor=MyDarkBlue, urlcolor=MyLightBlue}

\begin{document}
\vspace*{-0.2in}
\begin{flushright}
OSU-HEP-19-01
\end{flushright}
\vspace{0.5cm}
\begin{center}
{\Large\bf 
Simplest Radiative Dirac Neutrino Mass Models
}\\
\end{center}

\vspace{0.5cm}
\renewcommand{\thefootnote}{\fnsymbol{footnote}}
\begin{center}
{\large
{}~\textbf{Shaikh Saad}\footnote{ E-mail: \textcolor{MyLightBlue}{shaikh.saad@okstate.edu}}
}
\vspace{0.5cm}

{\em Department of Physics, Oklahoma State University, Stillwater, OK 74078, USA }
\end{center}

\renewcommand{\thefootnote}{\arabic{footnote}}
\setcounter{footnote}{0}
\thispagestyle{empty}

\begin{abstract}
If neutrinos are Dirac particles, their right-handed partners must be present in the theory. Once introduced in the Standard Model (SM), the difference between the baryon  number $B$ and the lepton number $L$ can be promoted to a local $U(1)_{B-L}$ symmetry since the corresponding gauge anomalies can be canceled by the right-handed neutrinos. Furthermore, the extremely small neutrino mass can be explained naturally if it is originated from the quantum correction. In this work, we propose simplest models of radiative Dirac neutrino mass using only $U(1)_{B-L}$ symmetry, and  without introducing additional fermions and without imposing ad hoc symmetries. In this simple framework, we provide  minimal models where Dirac neutrino mass appears at the  (i) one-loop, (ii) two-loop and (iii) three-loop. By performing systematic analysis, we show that the minimal one-loop model requires three beyond SM scalar multiplets, whereas minimal two-loop and three-loop  models require five. The presented  two-loop and three-loop Dirac mass models have not appeared in the literature before.     
\end{abstract}

\newpage
\setcounter{footnote}{0}
\section{Introduction}\label{SEC-01}
Neutrinos are massless in the Standard Model (SM), however, experimentally \cite{ Whitehead:2016xud, Decowski:2016axc, Abe:2017uxa, deSalas:2017kay} it is well established now that this cannot be true. Hence, the SM needs to be extended to incorporate neutrino mass. There exist  vast literature on neutrino mass generation, among them, radiative
\footnote{The first radiative neutrino mass model was proposed for Dirac neutrinos \cite{Cheng:1977ir}.} 
generation of neutrino mass is one of the most widely studied scenario. 
For models belonging to this class, two most popular mechanisms are: 
 the one-loop Zee-mechanism 
\cite{Zee:1980ai} 
and the two-loop Zee-Babu mechanism
\cite{Cheng:1980qt,Babu:1988ki}. 
In addition to yet unknown mechansim behind the neutrino mass generation, another greatest  mystery  of particle physics is whether the neutrino is Dirac or Majorana like particle. There is no theoretical preference and the issue needs to be settled experimentally. Great amount of experimental effort \cite{GERDA:2018zzh, KamLAND-Zen:2016pfg, Agostini:2017iyd, Kaulard:1998rb} is going on to come to a possible settlement.  However, since all the charged fermions in the SM acquire Dirac masses, it is quite natural to expect  that neutrino is also Dirac in nature.  
Theoretically the nature of the neutrinos may be closely related to the $U(1)_{B-L}$ symmetry and the way it breaks down, for details see Refs. \cite{Heeck:2013rpa, Hirsch:2017col}.   
Despite the vast literature on neutrino mass generation of the Majorana type
\footnote{
For a recent review on Majorana mass models see Ref. \cite{Cai:2017jrq}.},
 just recently Dirac neutrinos have gained a lot of attention to the particle physics community \cite{Ma:2014qra, Ma:2015mjd, Bonilla:2016zef, Chulia:2016giq, Bonilla:2016diq, Ma:2016mwh, Wang:2016lve, Chulia:2016ngi, Wang:2017mcy, Borah:2017dmk, Yao:2018ekp, Reig:2018mdk, Han:2018zcn, Calle:2018ovc, Bonilla:2018ynb, Jana:2019mez}. 

If neutrinos are Dirac in nature, then the right-handed partners must be present into the theory. 
The main difficulty faced while constructing a radiative Dirac neutrino mass model is to introduce ad hoc symmetries  beyond the SM (BSM) to forbid both the tree-level Dirac mass and the Majorana mass terms. Mechanisms implemented to forbid  these terms not only introduce  complexity in the theory but also fail to explain the reason behind the presence of these symmetries in the first place. Most of the constructions in the literature use ad hoc discrete symmetries,
except only few recent works \cite{Wang:2017mcy, Calle:2018ovc, Bonilla:2018ynb} that are based on $U(1)_{B-L}$ symmetry 
\footnote{
For Dirac neutrino mass models based on $SU(2)_L\times SU(2)_R\times U(1)_{B-L}$ symmetry see Refs.  
\cite{Cheng:1977ir, Mohapatra:1987hh, Mohapatra:1987nx, Balakrishna:1988bn, Babu:1988yq, Borah:2016zbd, Borah:2016hqn, Borah:2017leo, Bolton:2019bou}, on $U(1)_R$ symmetry see Ref. \cite{Jana:2019mez},
  on
$U(1)_L\times U(1)_R$ symmetry  see Ref. \cite{Rajpoot:1990gy}  
and on $SU(3)_L\times U(1)_X$ symmetry see Refs. \cite{Valle:1983dk, Valle:2016kyz, Reig:2016ewy}.} 
\footnote{For works relating Dirac neutrino mass and matter-antimatter asymmetry see for example Refs. \cite{Dick:1999je, Murayama:2002je, Gu:2006dc, Gu:2007ug, Gu:2012fg, Gu:2016hxh, Gu:2017bdw}.
}.
  However, in these works, even though $U(1)_{B-L}$  is the only symmetry used to forbid the unwanted terms,  extra fermion multiplets BSM are introduced to make the theory anomaly free, to  generate Dirac neutrino mass, and to have dark matter (DM) candidate. For most of the cases, new fermions come with more than one generation making these models less economical.  Furthermore, the presence of several  BSM scalar multiplets are also necessary to complete the loop diagrams.

Instead, to reduce the complexities in the theory, we aim to search for  the simplest  models of Dirac neutrino mass without the presence of the additional fermions.  Even though in this work we are not interested in the ultraviolet (UV) completion, but the motivation behind this  can be justified as follows.   The well motivated grand unified theory (GUT) based on $SO(10)$ gauge group, in its minimal realization contains only the SM fermions and the right-handed neutrinos. The other widely studied GUT which is based on $SU(5)$ gauge group  contains no extra fermions \footnote{In the context of renormalizable  $SU(5)$ GUT, different possibilities of generating Majorana neutrino mass are summarized in a recent work   \cite{Saad:2019vjo}
including new proposal of a  two-loop model.} beyond the SM, however, right-handed neutrinos can be trivially added since they are the gauge singlets.  
In this work we explore the possibilities of building 
 Dirac neutrino mass models using the baryon number minus the lepton number $U(1)_{B-L}$ symmetry offered by the presence of the right-handed neutrinos in the theory to form Dirac pair. 
Our set-up is the simplest in the sense that many of the complexities can be avoided  because,  
no additional fermions are introduced and no ad hoc symmetries are imposed to generate radiative Dirac neutrino mass. Within this framework, we propose minimal models of Dirac neutrino mass at the (i) one-loop, (ii) two-loop and (iii) three-loop levels. 
 By performing systematic analysis, we show that within the minimal scenario, one-loop realization  requires three BSM scalar multiplets  whereas two-loop and three-loop realizations require five multiplets. To the best of the author's knowledge, the presented two-loop and three-loop Dirac neutrino mass models in this work are not realized in the literature before. Though the presence of DM  is not demanded  in this set-up, but the possible DM candidate in this class of models is briefly discussed. 
 
In short, the main novelty of our work is to generate radiative Dirac neutrino mass with only $U(1)_{B-L}$ symmetry, no arbitrary  symmetries are imposed by hand and furthermore, other than the right-handed neutrinos that are required to make $U(1)_{B-L}$ anomaly free, no extra fermions are added to the SM.

\section{Framework}\label{SEC-02}
The SM has  accidental baryon number $U(1)_B$ and  lepton number  $U(1)_L$  conservation symmetries that are anomalous and cannot be gauged without the addition of quite a few number of new fermions into the theory. However, the difference between the  baryon number and the lepton number symmetry $U(1)_{B-L}$  can be made non-anomalous by adding only the right-handed partners, ${\nu_R}_i$ of the SM neutrinos, ${\nu_L}_i$.  Anomaly cancellation demands one of the two solutions, the lepton number of the right-handed neutrinos to be vector charges: $L_{1,2,3}=\{-1,-1,-1\}$  or   chiral charges: $L_{1,2,3}=\{+5,-4,-4\}$     under $U(1)_{B-L}$ \cite{Montero:2007cd, Machado:2010ui, Machado:2013oza}. Since all the charged fermions in the SM acquire Dirac masses, it is quite natural to expect that $\nu_L$ and $\nu_R$ pair up to form four-component Dirac fermion and  as a result the neutrinos also receive Dirac masses.

Since the $U(1)_{B-L}$ symmetry is anomaly free, it is natural to gauge this symmetry. Furthermore,  unlike discrete and global symmetries, gauge symmetries are known to be respected by  gravitational interactions. This is why we prefer to gauge $U(1)_{B-L}$, however, our analysis remains valid for both global and gauged $U(1)_{B-L}$.

Now, if the solution $L_{1,2,3}=\{-1,-1,-1\}$ as aforementioned is realized, then just like all the charged fermions, neutrinos receive Dirac masses  at the tree-level when the Electroweak (EW) symmetry is broken by the SM Higgs VEV from the gauge invariant Yukawa term: $\mathcal{L}_Y\supset y^{\nu}\overline{L}_L\nu_R \widetilde{H}$,  where $L_L=(\nu_L\; \ell_L^-)^T$, $H=(H^+\; H^0)^T$ and $\widetilde{H}=\epsilon H^{\ast}$ ($\epsilon$ is the 2-index Levicivita tensor). However, this tree-level Dirac mass realization requires the corresponding Yukawa couplings to be of the order of $y^{\nu}\sim 10^{-11}$. This may not be a natural solution. However, small Dirac mass for neutrinos can be  realized naturally  if the second  anomaly free solution  $L_{1,2,3}=\{+5,-4,-4\}$ is considered \cite{Ma:2014qra}, this choice automatically forbids the tree-level   Dirac mass term. In our framework, we adopt this chiral charge solution, hence tree-level Dirac mass is forbidden and neutrinos receive  Dirac masses from the quantum corrections, which naturally explains the smallness of the observed neutrino mass. Furthermore, we do not require ad hoc symmetries  to forbid the  Majorana mass terms for $\nu_L$ and $\nu_R$, hence our set-up is  theoretically well motivated. Within this framework, it is worthwhile to search for the minimal models to generate non-zero neutrino mass.  The new idea of this paper is to utilize  the chiral anomaly free solution and construct Dirac neutrino mass radiatively  without introducing new fermions and without imposing ad hoc symmetries.

In this work we construct  simple ways of neutrino mass generation by demanding the following requirements:  
\begin{itemize}

\item Neutrinos are Dirac type particles. 

\item Neutrino mass originates from quantum correction.

\item No additional fermions are introduced except the three right-handed neutrinos. 

\item No additional symmetries are imposed except the $U(1)_{B-L}$ that is automatically offered by the presence of the right-handed neutrinos.  

\end{itemize}

\begin{figure}[t!]
\begin{subfigure}[b]{.5\linewidth}
\centering\includegraphics[scale=0.35]{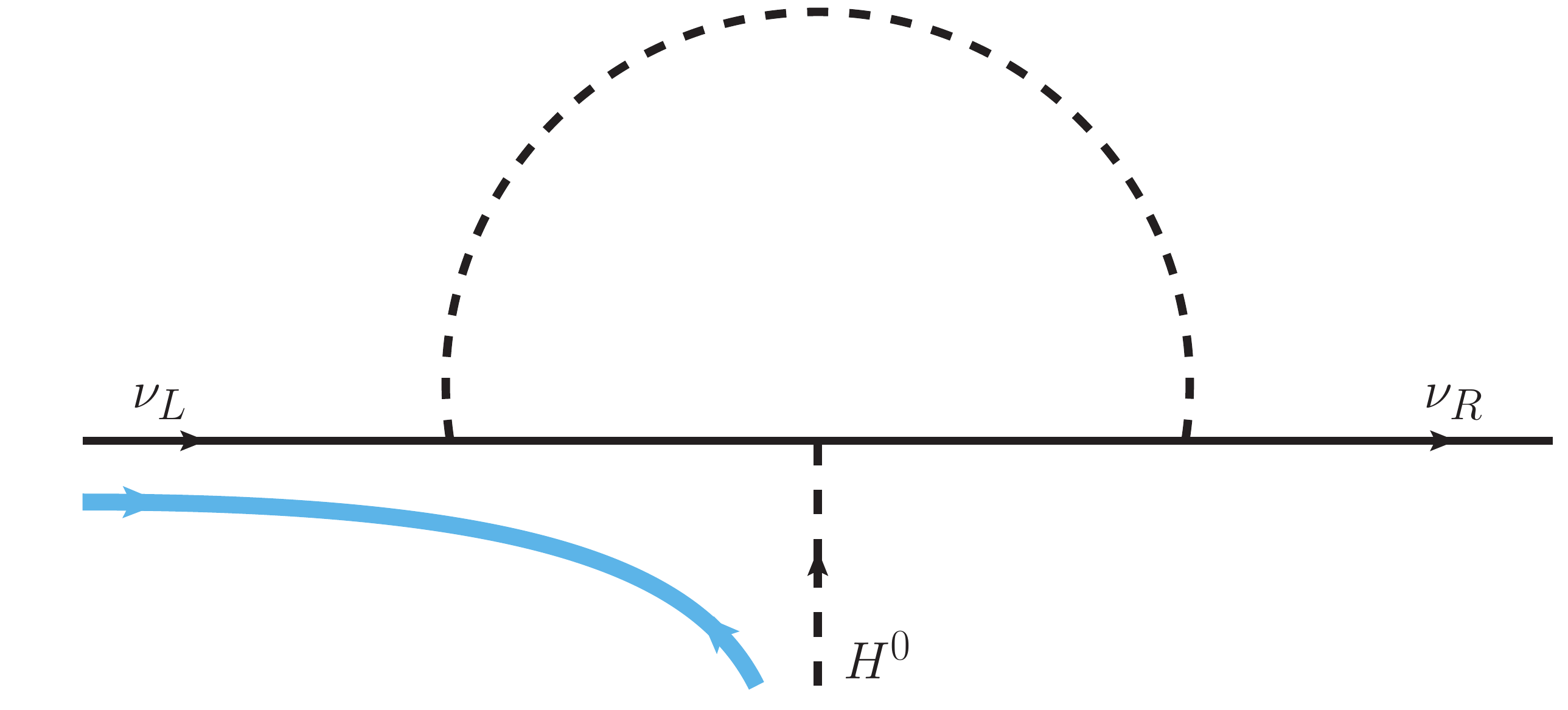}\hspace{0.2in}
\caption{T-I-F-i}
\end{subfigure}
\hspace{0.1in}
\begin{subfigure}[b]{.5\linewidth}
\centering\includegraphics[scale=0.35]{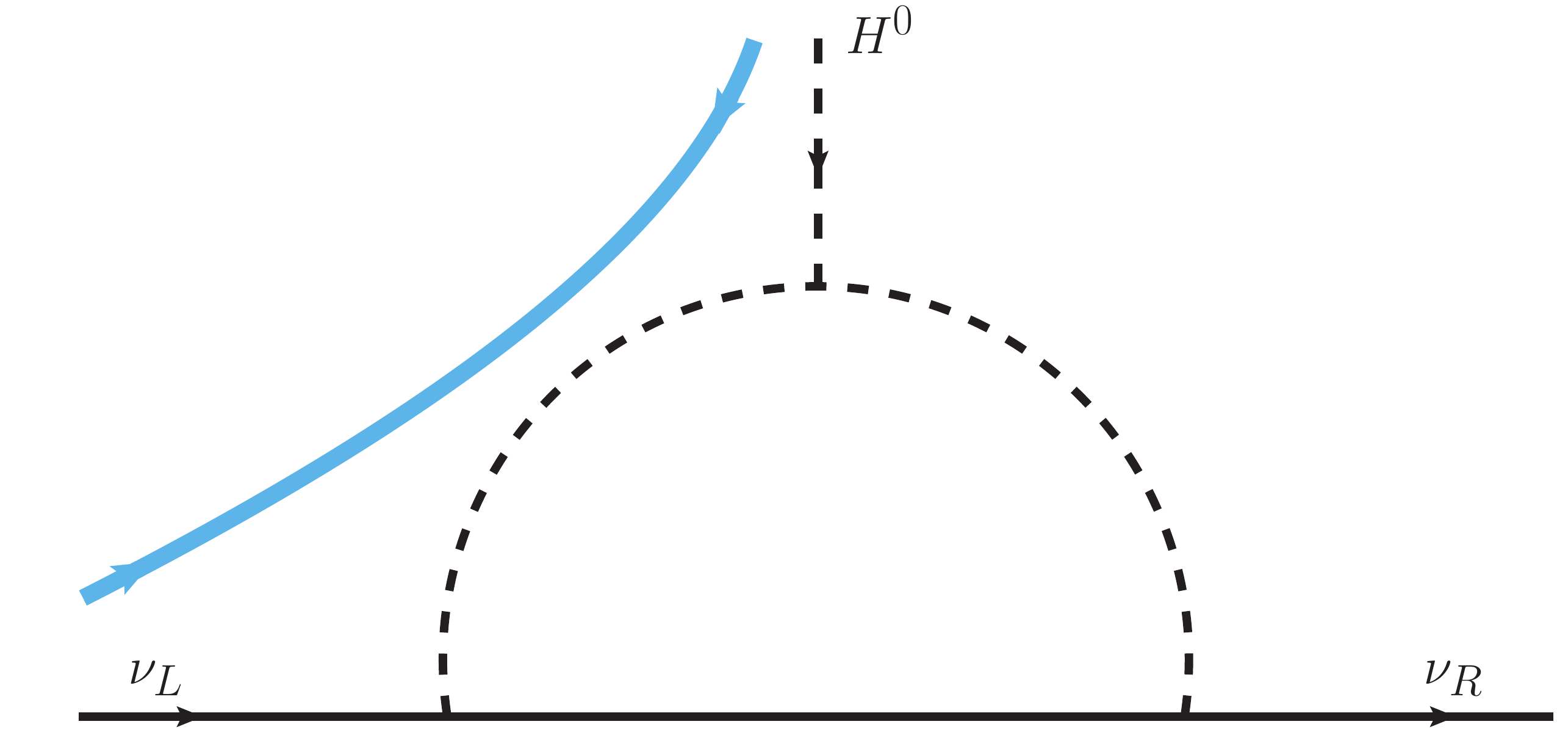}\hspace{0.2in}
\caption{T-I-S-i}
\end{subfigure}
\caption{ Expected one-loop topologies within our framework. The blue colored arrow represents the direction of the iso-spin doublet flow. The different labeling we use to differentiate among different topologies are  explained in the text. }\label{one-loop}
\end{figure}

\begin{figure}[b!]
\begin{subfigure}[b]{0.3\linewidth}
\centering\includegraphics[scale=0.35]{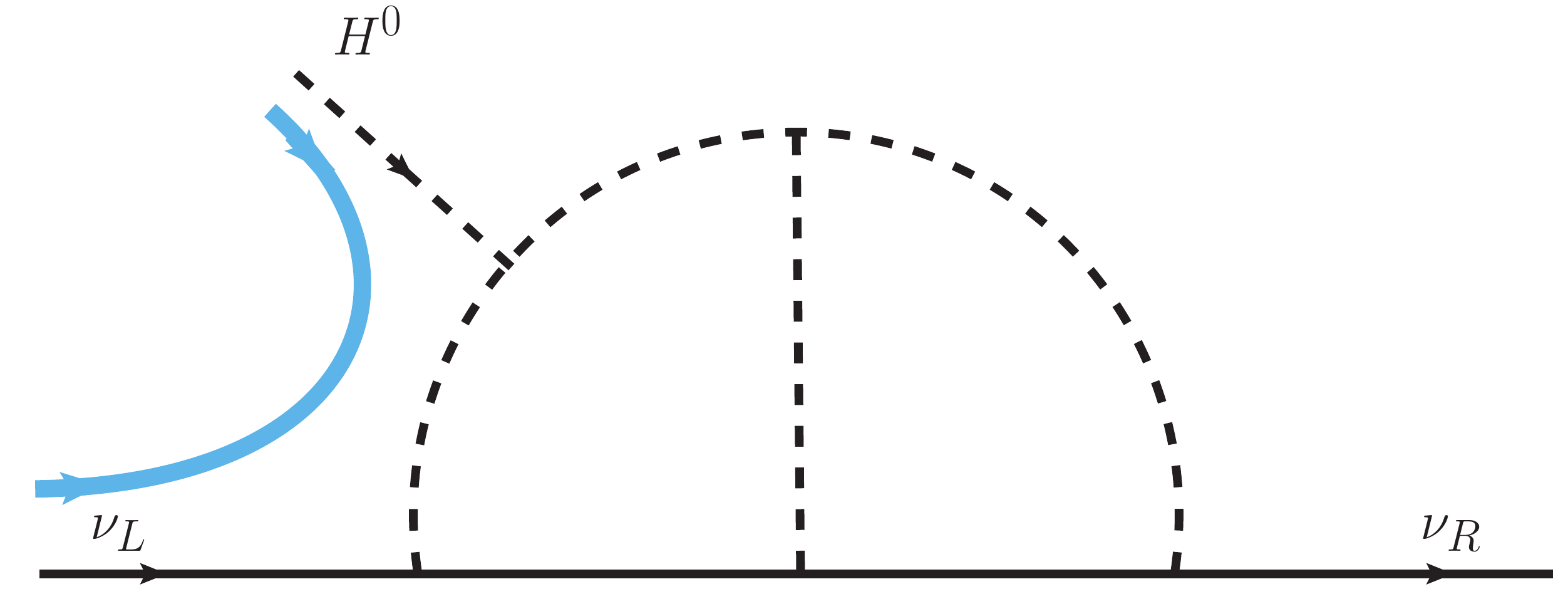}\hspace{0.2in}
\caption{T-II-S-i}
\end{subfigure}
\hspace{1.3in}
\begin{subfigure}[b]{0.3\linewidth}
\centering\includegraphics[scale=0.35]{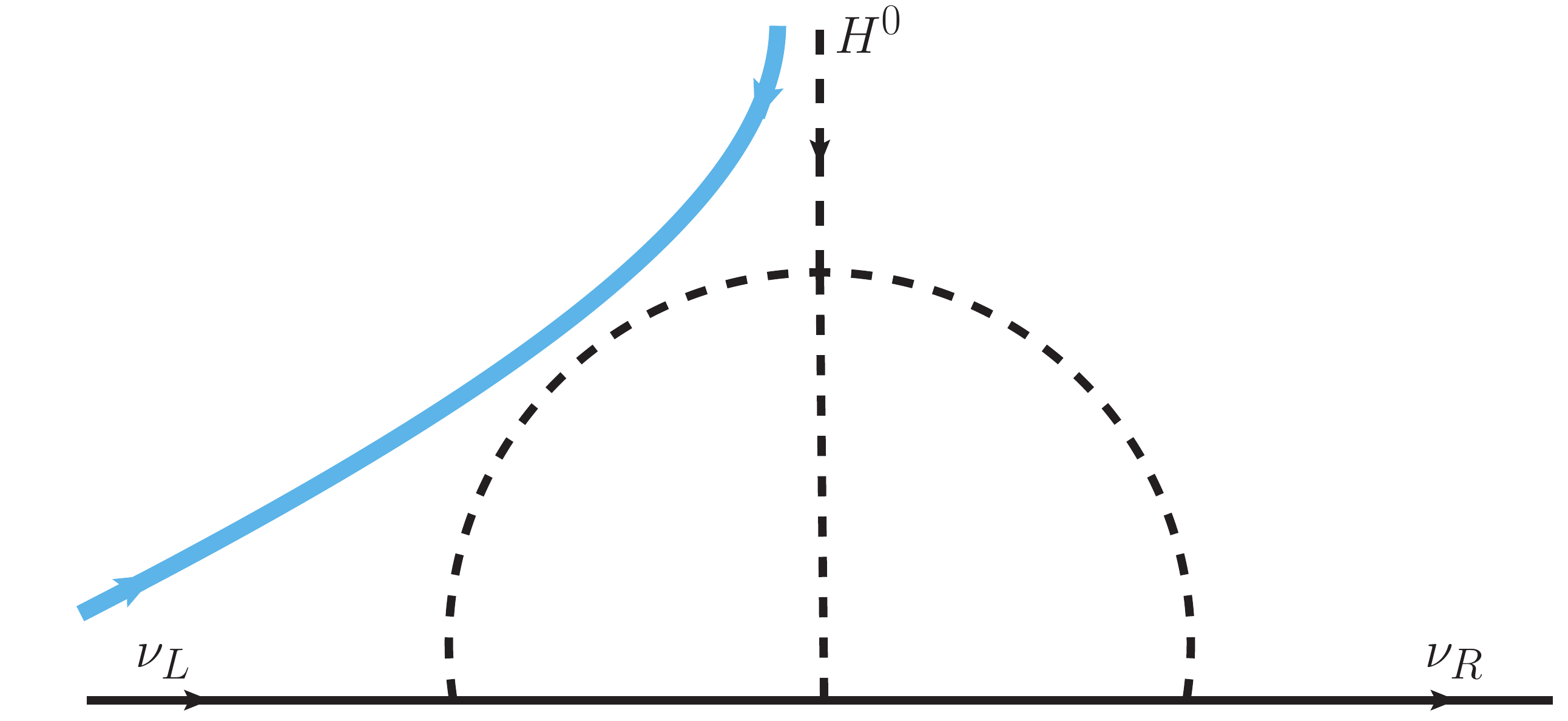}
\caption{T-II-S-ii}
\end{subfigure}
\vspace{0.5in}\\ 
\begin{subfigure}[b]{.3\linewidth}
\centering\includegraphics[scale=0.35]{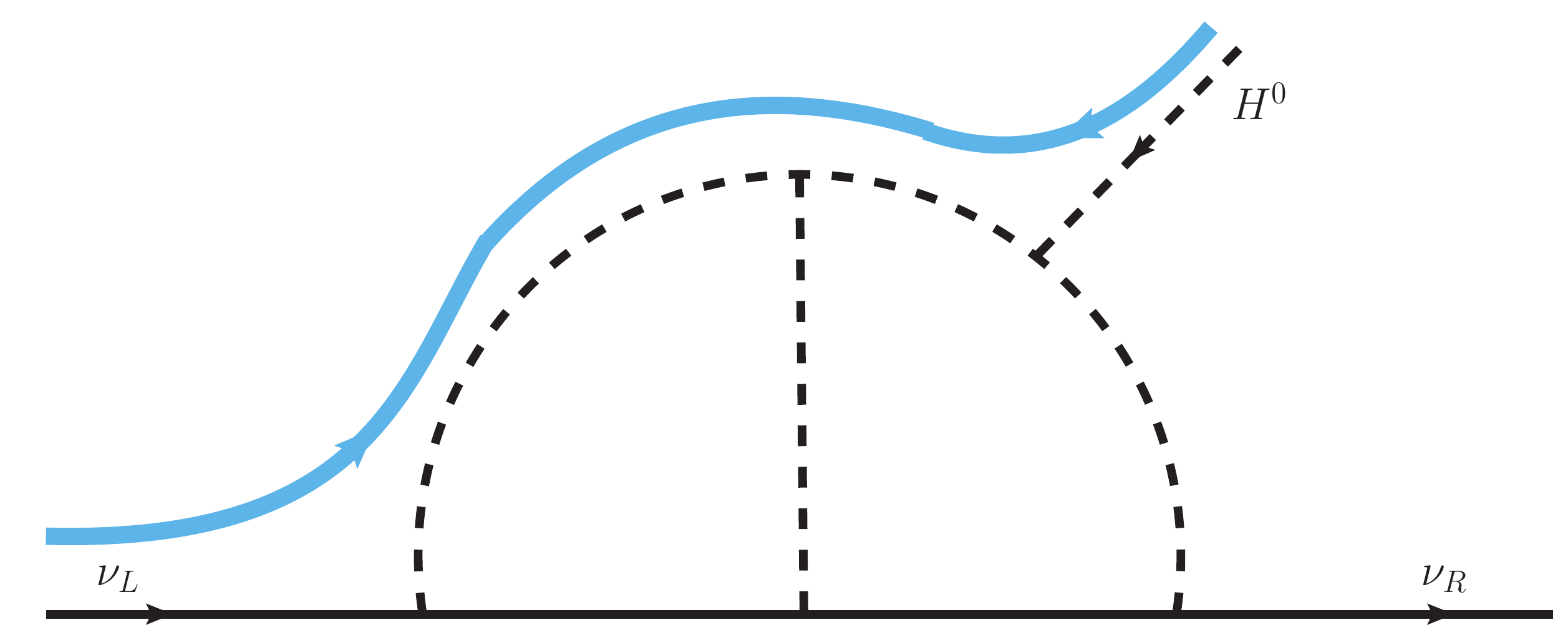}\hspace{0.2in}
\caption{T-II-S-iii}
\end{subfigure}
\hspace{1.3in}
\begin{subfigure}[b]{.3\linewidth}
\centering\includegraphics[scale=0.35]{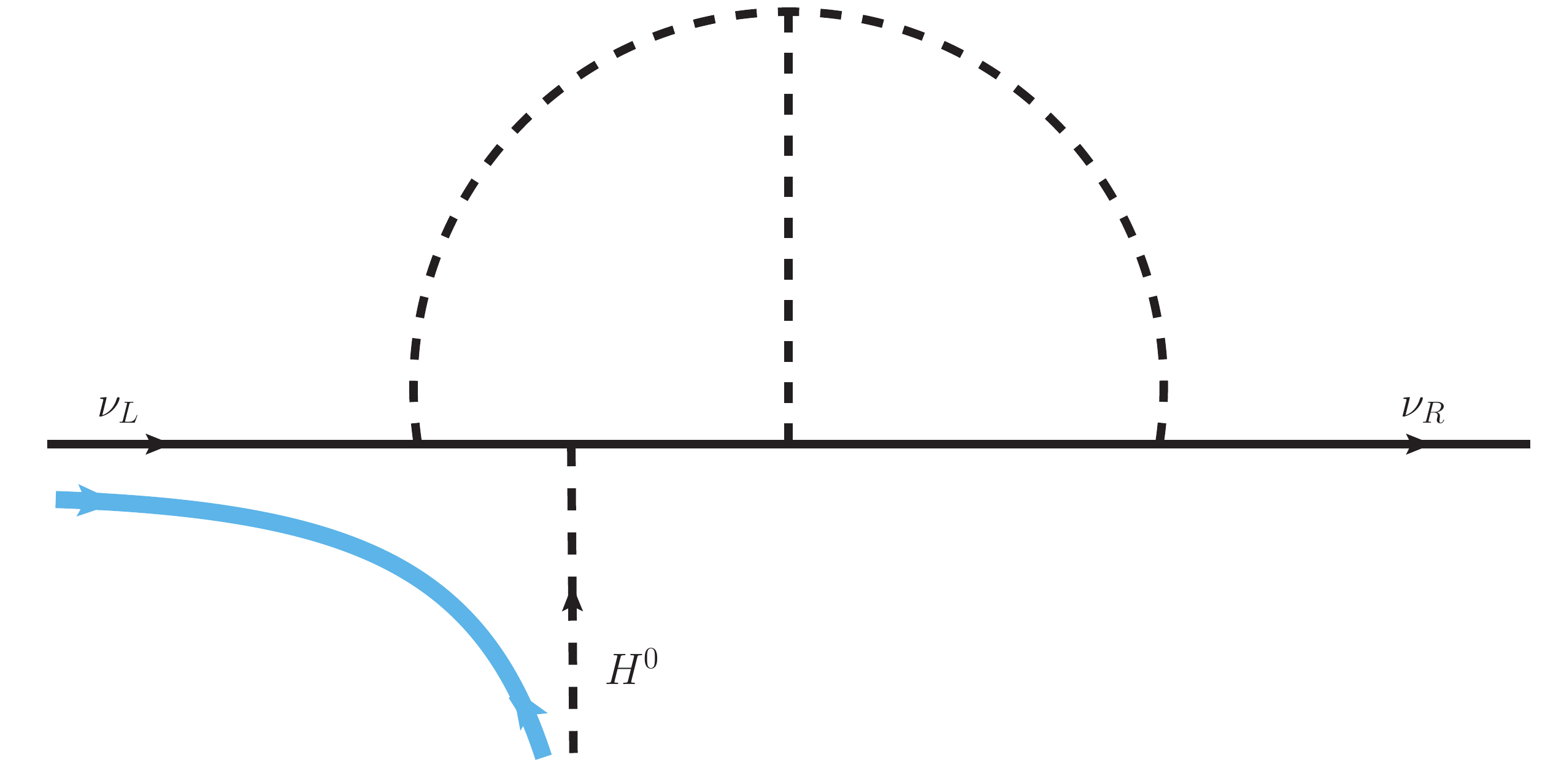}
\caption{T-II-F-i}
\end{subfigure}
\caption{ Expected two-loop topologies within our framework. The blue colored arrow represents the direction of the iso-spin doublet flow. The different labeling we use to differentiate among different topologies are explained in the text. }\label{two-loop}
\end{figure}

\noindent
We provide examples of minimal models for one-loop, two-loop and three-loop scenarios that obey the requirements listed above.   In our framework, the lowest order operator to generate radiative Dirac neutrino mass is the following  dimension-5 operator:     
\begin{align}
\mathcal{L}_5= \frac{h_{ij}}{\Lambda}\;\overline{L_L}_i \widetilde{H}\;{\nu_R}_j\sigma^{0}. \label{operator}
\end{align}

\noindent
Here the neutral scalar $\sigma^0$ is a  singlet under the  SM and 
we assume that our theory allows the presence of this operator.
 Based on this dimension-5 operator we build minimal models of  one-loop and two-loop Dirac mass for neutrinos.   We will show that 
it is possible to construct three-loop Dirac mass models by allowing this lowest dimensional operator, however, minimal model can be constructed by utilizing  dimension-7 operator given by:    
\begin{align}
\mathcal{L}_7= \frac{h_{ij}}{\Lambda^3}\;\overline{L_L}_i \widetilde{H}\;{\nu_R}_j\sigma^{0} (H^{\dagger}H). \label{operator-7}
\end{align}

From Eq. \eqref{operator}, the topologies that  we are interested in  can be constructed  straightforwardly and are presented in Fig. \ref{one-loop} for one-loop scenario and in Fig. \ref{two-loop} for the case of two-loop. We will use these topologies as a guidance to build minimal models within our set-up.  We stress the fact that not all the topologies can satisfy the set of  requirements listed above and we will discuss them case by case.  In these figures, the blue colored arrow represents the direction of the iso-spin doublet flow. Each of the different topologies is labeled as: T-x-y-z, where x=I for one-loop (x=II for two-loop), y=F corresponds to the case when the SM Higgs doublet is attached to fermion line (y=S when the SM Higgs doublet is attached to scalar line) and z=i, ii, ... to differentiate among topologies with the same x and y labels. 
The topologies corresponding to the three-loop models will be presented later in the text.

\FloatBarrier
\begin{table}[th!]
\centering
\footnotesize
\resizebox{0.5\textwidth}{!}{
\begin{tabular}{|c|c|c|}
\hline
Multiplets& $SU(2)_L\times U(1)_Y\times U(1)_{B-L}$   \\ \hline\hline
Leptons&
\pbox{10cm}{
\vspace{2pt}
${L_L}_i (2,-\frac{1}{2},\textcolor{blue}{-1})$\\
${\ell_R}_i (1,-1,\textcolor{blue}{-1})$\\
${\nu_R}_i (1,0,\{\textcolor{blue}{5,-4,-4}\})$
\vspace{2pt}}
\\ \hline\hline
Higgs &
\pbox{10cm}{
\vspace{2pt}
$H (2,\frac{1}{2},\textcolor{blue}{0})$
\vspace{2pt}}
\\ \hline
\end{tabular}
}
\caption{ 
Quantum numbers of the leptons and SM Higgs doublet.
}\label{charge-SM}
\end{table}

First we would like to determine the BSM scalar multiplets to complete these loop diagrams. 
To do so, first we look at the structures of the possible  fermion bilinears.  
These will also revel the multiplets that can give rise to the unwanted tree-level Dirac or Majorana mass terms. We only introduce BSM scalars that are color blind, hence the symmetry that we are interested in is: $SU(2)_L\times U(1)_Y\times U(1)_{B-L}$. Under this symmetry, the quantum numbers of the leptons and the SM Higgs doublet are presented in Table \ref{charge-SM}.  Since we do not introduce additional fermions, the number of possible fermion bilinears are limited and are given by:
\begin{itemize}

\item $\overline{L_L}\otimes\ell_R \sim (2,-\frac{1}{2},0)$

\item $\overline{L^c_L}\otimes L_L \sim (1,-1,-2) \oplus \color{red}(3,-1,-2)\color{black}$

\item $\overline{L_L}\otimes\nu_R \sim \color{red}
(2,\frac{1}{2},+6)
\color{black} 
\oplus 
\color{red}
(2,\frac{1}{2},-3)$
\color{black}

\item $\overline{\ell^c_R}\otimes\ell_R \sim (1,-2,-2)$

\item  $\overline{\ell^c_R}\otimes\nu_R \sim 
(1,-1,+4) 
\oplus 
(1,-1,-5)$ 

\item  $\overline{\nu^c_R}\otimes\nu_R \sim 
\color{red}
(1,0,+10) 
\color{black} 
\oplus 
\color{red}
(1,0,+1) 
\color{black} 
\oplus 
\color{red}
(1,0,-8)$
\color{black} 
\end{itemize}

\noindent
From these bilinears, it is clear that to avoid tree-level neutrino Dirac mass term, 
$\color{red}
(2,-\frac{1}{2},-6)
\color{black}$
 and  
$\color{red}
(2,-\frac{1}{2},+3)$
Higgs multiplets cannot be present. If they do, one needs to make sure that the electrically neutral component of the corresponding doublet field does not acquire any VEV, which might be a difficult task to arrange and may require the existence of some additional discrete symmetries. Furthermore, the iso-spin triplet Higgs  $\color{red}(3,+1,+2)\color{black}$ can be problematic if it is allowed to acquire VEV which will lead to tree-level type-II seasaw Majorana mass for the left-handed neutrinos.  And similarly, to forbid the tree-level Majorana mass terms for the right-handed neutrinos 
\color{red}
$(1,0,-10)$\color{black},   
\color{red}
$(1,0,-1)$ 
\color{black} 
 and
\color{red}
$(1,0,+8)$
\color{black} 
Higgs multiplets are not allowed in our theory if they were to acquire VEVs.
 
For our model building purpose, we use the structure of the above bilinears to select the appropriate Higgs multiplets for radiative neutrino mass generation. So the $U(1)_{B-L}$ symmetry plays the crucial role to forbid all the unwanted terms.  As long as this symmetry is preserved, neutrinos are massless at all order of the perturbation theory within our framework. However, due to spontaneously  breaking of the  $U(1)_{B-L}$ symmetry, neutrinos receive Dirac mass via quantum correction.   We achieve this symmetry breaking by the SM singlet scalar $\sigma^0$ that carries a non-trivial charge under $U(1)_{B-L}$. 
When this field acquires a VEV, the dimension-5 operator of Eq. \eqref{operator} can in principle generate Dirac neutrino mass within our set-up at the one-loop, two-loop and at the three-loop depending on the presence of certain BSM Higgs multiplets. We construct different possible UV completions of this lowest dimensional operator in search of  minimal models at the one-loop, two-loop and three-loop. 
Even though we can successfully construct minimal models for one-loop and two-loop scenarios, implementation of the dimension-7 operator of Eq. \eqref{operator-7} is more economical for the  three-loop scenario to be discussed in great details later in the text.

As a result of the $U(1)_{B-L}$ symmetry breaking, the imaginary part of $\sigma^0$ will  be eaten up by the  corresponding gauge boson $Z^{\prime}$. In this framework $Z^{\prime}$ can be made sufficiently heavy to evade collider constraints. 
For low $Z^{\prime}$ mass, interesting phenomenology may emerge, however in this work we do not discuss such possibilities rather focus mainly on the ways of neutrino mass generation. For detail phenomenologies associated with  $Z^{\prime}$ arising from different $U(1)$ models we refer the readers to this review \cite{Langacker:2008yv}.

Based on this simple framework discussed in details in this section, we construct the corresponding minimal models at the one-loop, two-loop and three-loop in Secs. \ref{one}, \ref{two} and \ref{three} respectively. We also analyze the effects of the higher dimensional operators of these models in appendices \ref{A},  \ref{B} and  \ref{C}. 

\section{Minimal one-loop model}\label{one}
For one-loop Dirac neutrino mass generation, there are two different expected topologies depending on the direction of the  flow of the iso-spin doublet  within our framework as shown in Fig. \ref{one-loop}. However, between these two possibilities, model with  topology T-I-F-x can only be realized. The absence of the T-I-S-x  topology is due to
not allowing additional fermions BSM, which is one of our requirements.  
As a result the internal fermion line cannot be completed  for the diagrams of type  T-I-S-x. 
However, by relaxing this requirement, topology of this type can be built  by utilizing  $U(1)_{B-L}$ symmetry  \cite{Wang:2017mcy, Calle:2018ovc, Bonilla:2018ynb}.

\FloatBarrier
\begin{table}[th!]
\centering
\footnotesize
\resizebox{0.5\textwidth}{!}{
\begin{tabular}{|c|c|}
\hline
Topology &$SU(2)_L\times U(1)_Y\times U(1)_{B-L}$  \\ \hline\hline
T-I-F-i&
\pbox{10cm}{
\vspace{2pt}
$\sigma^0(1,0,\textcolor{blue}{3})$\\
$S^+(1,1,\textcolor{blue}{5})$\\
$\xi^+(1,1,\textcolor{blue}{2})$
\vspace{2pt}}
\\ \hline
\end{tabular}
}
\caption{ 
Quantum numbers of the BSM scalars.
}\label{tab:T-I-F-i}
\end{table}
\FloatBarrier
\begin{figure}[th!]
\centering
\includegraphics[scale=0.4]{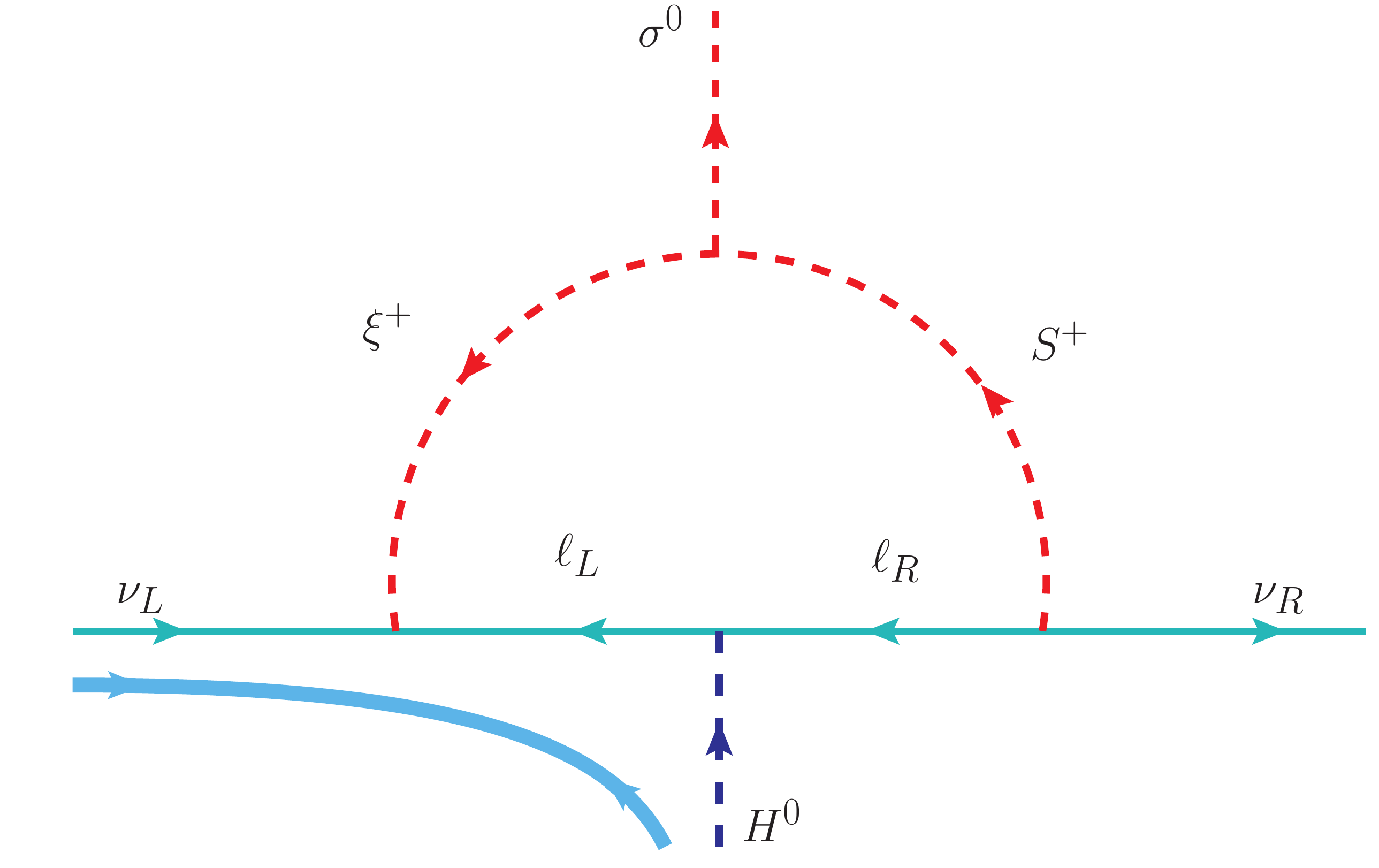}
\caption{One-loop Dirac neutrino mass for the particle content shown in Table \ref{tab:T-I-F-i}. Propagators in red correspond to BSM scalars. }\label{fig:T-I-F-i}
\end{figure}   

A very simple model with T-I-F-x  topology can be constructed by extending the SM Higgs sector by only three scalars: a SM singlet scalar to break the $U(1)_{B-L}$ symmetry and two singly charged scalars.  
The quantum numbers of these multiplets are presented in Table \ref{tab:T-I-F-i} and the corresponding Feynman diagram is shown in Fig. \ref{fig:T-I-F-i}.  
This is the most minimal model of radiative Dirac neutrino mass existing in the literature.  The simplicity arises as a result of not imposing any  ad hoc symmetry, no BSM fermion is  introduced and in this case only three BSM Higgs multiplets are required. Any other  model demands more than one BSM chiral fermions along with more than one BSM scalar multiplets, making the model more complicated. By introducing a $\mathcal{Z}_2$ symmetry, this diagram was first  realized in \cite{Nasri:2001ax} and later discussed in \cite{Kanemura:2011jj}. In the context of  $U(1)_{B-L}$ symmetry, this model was  just recently reappeared  in \cite{Calle:2018ovc}.

In this model the relevant Yukawa Lagrangian is given by:
\begin{align}
\mathcal{L}_Y \supset
y^{H}_{ij}\overline{L_L}_i{\ell_R}_jH
+
y^{S}_{kj}{\overline{\nu_R^c}}_k{\ell_R}_jS^+
+
y^{\xi}_{ij}\overline{L_L^c}_i\epsilon{L_L}_j\xi^+
+h.c.
\end{align}

\noindent Where $i,j=1-3$ and $k=2,3$ are the generation indices. Note that $y^{\xi}$ is a $3\times 3$ anti-symmetric matrix, and since ${\nu_R}_1$ carries a different $U(1)_{B-L}$ charge compared to the other two generations as a result 
 one has  $y^{S}_{11,12,13}=0$.
The relevant part of the scalar potential to complete the loop diagram contains the following cubic term:
\begin{align}
V\supset
\mu\;S^+\sigma^{0\ast}\xi^- 
+h.c.
\end{align}

Neutrino mass will be generated when the  neutral singlet $\sigma^0$ that carries $+3$ unit of charge under the $U(1)_{B-L}$ acquires  non-zero VEV. EW symmetry also needs to be broken for this diagram to contribute to neutrino mass.  This specific charge assignment of $\sigma^0$ multiplet is uniquely fixed by the associated topology  \ref{fig:T-I-F-i}.  
The corresponding neutrino mass matrix has the form:
\begin{align} \label{numass-1}
{m_{\nu}}_{ab}\sim \frac{1}{16\pi^2}
\frac{\mu\langle \sigma^0 \rangle \langle H^0 \rangle}{\Lambda^2}\;
y^{\xi}_{ai}\;y^H_{ij}\;y^{S}_{jb}.
\end{align} 

\noindent
Here $\Lambda$ is the mass scale associated with the heaviest BSM scalar multiplet running in the loop. 
The Yukawa coupling $y^{\xi}$ is anti-symmetric, as a result, one of the neutrinos remains massless to all order since $det(y^{\xi})=0$, irrespective of the restrictions on $y^S$.   Since ${\nu_R}_1$ has no interaction due to a different $B-L$ charge assignment, it is one of the two massless chiral states and  can in principle 
 contribute to  $N_{eff}$, the effective number of relativistic degrees of freedom.  However, this chiral state does not couple directly to the SM particles, and it decouples from the rest of the theory early in the universe.
As a result  ${\nu_R}_1$ is not necessarily produced and may not contribute to $N_{eff}$ at all, 
 hence does not conflict with any cosmological measurement  \cite{Weinberg:2013kea, Perez:2017qns, Nomura:2017jxb, Han:2018zcn}.

After the breaking of the $U(1)_{B-L}$ and EW symmetries, the Dirac neutrino mass term in the Lagrangian has the form $\mathcal{L}_Y\supset \overline{\nu}_L m_{\nu} \nu_R$. One can go to a basis where the charged lepton mass matrix is diagonal. The phases of the entries $y^{\xi}_{ij}$ can be absorbed in $L_L$ (and subsequently in $\ell_R$). Then $y^S$ is in general complex, however, recall that $y^S_{1j}=0$. Since neutrinos are Dirac particles, one can also simultaneously work in a basis where ${\nu_R}_i$ are mass eigenstates. Then the Dirac mass matrix can be written as: $m_{\nu}=U m_{\nu}^{diag}$, with $U=U_{PMNS}$. Assuming normal mass ordering, this corresponds to:
\begin{align}
m_{\nu}=a_0{y^{\xi}}^Tm_E\;y^S
=\begin{pmatrix}
0&m_2U_{e2}&m_3U_{e3}\\
0&m_2U_{\mu 2}&m_3U_{\mu 3}\\
0&m_2U_{\tau 2}&m_3U_{\tau 3}
\end{pmatrix},\;\;\;
a_0=\frac{\sin(2\theta)}{16\pi^2}\text{ln}\left( \frac{m^2_{H_2}}{m^2_{H_1}} \right).
\end{align} 

\noindent Where, $H_i$ ($i=1,2$) represents the singly  charged scalar mass eigenstate and $\theta$ represents the corresponding mixing angle. Solving this system fixes the Yukawa couplings as ($y^S_{21,31}$ remains undetermined):
\begin{align}
&y^S_{22}=\frac{m_2}{m_{\mu}}\left(\frac{-c_{23}s_{12}s_{13}-c_{12}s_{23}}{a_0\;y^{\xi}_{23}}\right), 
\;\;
y^S_{23}=\frac{m_3}{m_{\mu}}\left(\frac{c_{13}c_{23}}{a_0\;y^{\xi}_{23}}\right),
\\
&y^S_{32}=\frac{m_2}{m_{\tau}}\left(\frac{s_{12}s_{13}s_{23}-c_{12}c_{23}}{a_0\;y^{\xi}_{23}}\right),
\;\;
y^S_{33}=\frac{m_3}{m_{\tau}}\left(\frac{-c_{13}s_{23}}{a_0\;y^{\xi}_{23}}\right),
\\
&y^{\xi}_{12}=y^{\xi}_{23} \left(\frac{-c_{12}c_{23}s_{13}+s_{12}s_{23}}{s_{13}}\right),
\;\;
y^{\xi}_{13}=y^{\xi}_{23} \left(\frac{-c_{23}s_{12}-c_{12}s_{13}s_{23}}{s_{13}}\right).
\end{align}

\noindent Here $c_{ij}=\cos(\theta^{PMNS}_{ij})$ and $s_{ij}=\sin(\theta^{PMNS}_{ij})$, $m_2=\sqrt{\Delta m^2_{21}}$  and $m_3=\sqrt{\Delta m^2_{31}}$. For simplicity, here we assumed $U_{PMNS}$ is real by setting the Dirac phase $\delta=0$. It is interesting to note that all these Yukawa couplings are fixed in terms of $y^{\xi}_{23}$ and by the neutrino oscillation data, hence parameter space of this model is highly restricted. These non-trivial restrictions can have important significance in phenomenology of this model.

\FloatBarrier
\begin{figure}[th!]
\centering\includegraphics[scale=0.45]{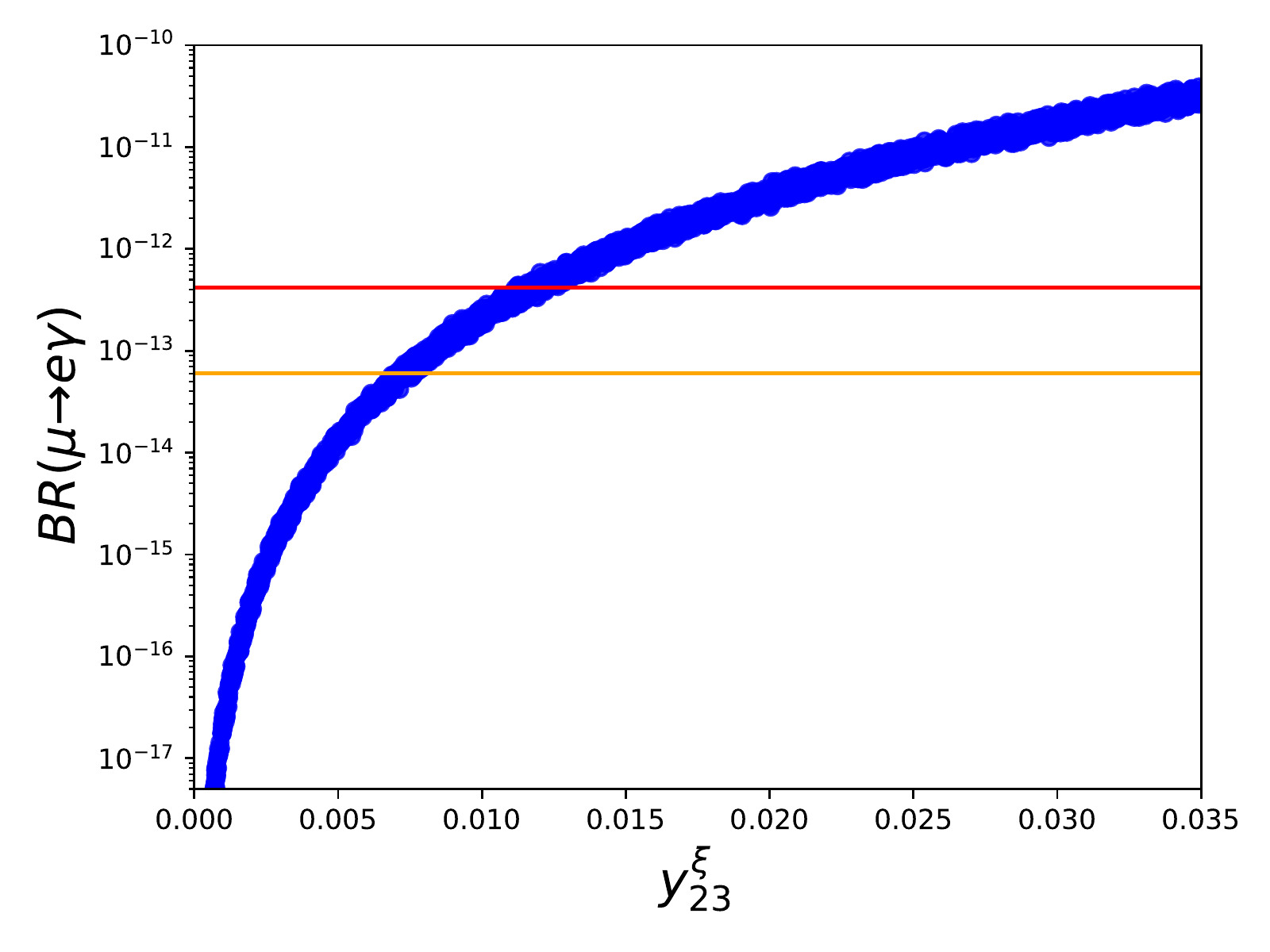}\hspace{0.2in}
\caption{In this figure we plot $BR(\mu\to e\gamma)$ as a function of $y^{\xi}_{23}$, for details see text. The  horizontal red line corresponds to the current experimental upper bound  $BR(\mu\to e\gamma)<4.2\times 10^{-13}$ by MEG experiment and the horizontal orange line represents the projected sensitivity $BR(\mu\to e\gamma)<6\times 10^{-14}$ by MEG II experiment  \cite{TheMEG:2016wtm}.  }\label{BR}
\end{figure}

The presence of the Yukawa couplings $y^{\xi}_{ij}$ give rise to lepton flavor violating (LFV) decays, $\ell_i \to \ell_j \gamma$, among them the most stringent constraints   comes from $\mu \to e \gamma$ process. Using the above derived relations, we find the corresponding branching ratio to be:
\begin{align}\label{brI}
BR(\mu\to e\gamma)=\frac{\alpha}{48\pi G^2_F}
A^2
\left(y^{\xi}_{23}\right)^4
\left( \frac{c_{23}s_{12}+c_{12}s_{13}s_{23}}{s_{13}} \right)^2,\;\;\;
A=\frac{\cos^2\theta}{m^2_{H_1}}+\frac{\sin^2\theta}{m^2_{H_2}}.
\end{align}  

\noindent 
Note that  the Yukawa coupling $y^S$ does not contribute to this process in our scenario due to $y^S_{1j}=0$.  
Assuming $\theta=0.1$ and setting $m_{H_{1,2}}=650, 750$ GeV, the $\mu\to e\gamma$ branching ratio is plotted against the only free Yukawa coupling, $y^{\xi}_{23}$ that  appears in Eq. \eqref{brI}  and shown in Fig. \ref{BR}. In this plot all the neutrino observables are varied within their current experimental $3\sigma$ range \cite{deSalas:2017kay}. Other interesting processes of this model are $\mu\to e\overline{e}e$ and $\mu - e$ conversion in the  nuclei and so on, 
for general analysis of  LVF processes and for LHC phenomenology  we refer the reader to  \cite{Nasri:2001ax, Kanemura:2011jj}.

\FloatBarrier
\begin{table}[b!]
\centering
\footnotesize
\resizebox{0.5\textwidth}{!}{
\begin{tabular}{|c|c|}
\hline
Topology &$SU(2)_L\times U(1)_Y\times U(1)_{B-L}$  \\ \hline\hline
T-II-S-i&
\pbox{10cm}{
\vspace{2pt}
$\sigma^0(1,0,\textcolor{blue}{3})$\\
$S^+(1,1,\textcolor{blue}{5})$\\
$\chi^{++}(1,2,\textcolor{blue}{2})$\\
$\eta(2,\frac{1}{2},\textcolor{blue}{0})$\\
$\Omega^+(1,1,\textcolor{blue}{-3})$
\vspace{2pt}}
\\ \hline
\end{tabular}
}
\caption{ 
Quantum numbers of the BSM scalars.
}\label{tab:T-II-S-i}
\end{table}
\begin{figure}[b!]
\centering
\includegraphics[scale=0.4]{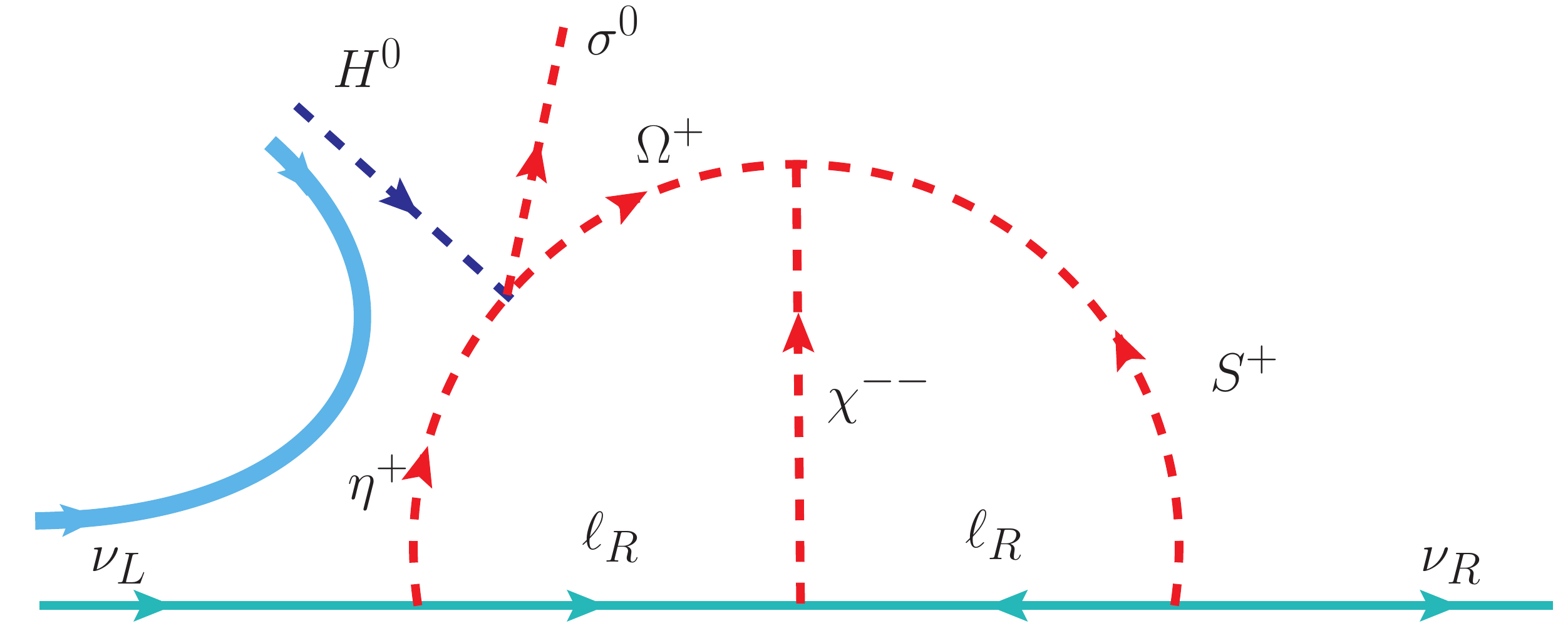}
\caption{Two-loop Dirac neutrino mass for the particle content shown in Table \ref{tab:T-II-S-i}.  }\label{fig:T-II-S-i}
\end{figure}   

\section{Minimal two-loop models}\label{two}
In this section, we present the simplest two-loop  Dirac neutrino mass models within our framework.  As aforementioned, we are interested in topologies presented in Fig. \ref{two-loop} as a realization of the dimension-5 operator of Eq. \eqref{operator}. 
First we point out that, out of the four topologies, the topology  of the type T-II-F-x   cannot be generated with only the SM fermions without inducing one-loop diagram (Fig. \ref{fig:T-I-F-i}), hence we discard this possibility.  Similar to that of  T-I-S-x topology as discussed in the previous section, the internal fermion line cannot be completed without the existence of   BSM chiral fermions to build a true two-loop model.  Below we present the minimal  models with topology of the  type T-II-S-x. We will show that realization of this topology in our set-up demands five  BSM scalar multiplets. 
 By comparison with the existing two-loop Dirac neutrino mass models in the literature, models presented in this work require less number of BSM states and simple in nature.

For topology T-II-S-i the
particle content with quantum numbers is presented in Table \ref{tab:T-II-S-i} and the Feynman diagram is shown in Fig. \ref{fig:T-II-S-i}. A second Higgs doublet neutral under $U(1)_{B-L}$ is required in this model. To complete the loop, in addition to two singly charged scalars, a doubly charged scalar is introduced, these three multiplets along with the neutral scalar that breaks the $U(1)_{B-L}$ symmetry are all iso-singlets and carry non-trivial $B-L$ charge.    
The relevant Yukawa couplings are given by:
\begin{align}
\mathcal{L}_Y \supset
y^{H}_{ij}\overline{L_L}_i{\ell_R}_jH
+
y^{\eta}_{ij}\overline{L_L}_i{\ell_R}_j\eta
+
y^{S}_{kj}\overline{\nu_R^c}_k{\ell_R}_jS^+
+
y^{\chi}_{ij}\overline{\ell_R^c}_i{\ell_R}_j\chi^{++}
+h.c.
\label{LY:T-II-S-i}
\end{align}

\noindent
Here the Yukawa coupling of the doubly charged scalar, $y^{\chi}$ is a $3\times 3$ symmetric matrix in the generation space. And the relevant part of the scalar potential that contributes to complete the loop diagram is:
\begin{align}
V\supset 
\mu\;S^+\Omega^+\chi^{--}
+
\lambda\;H\epsilon \eta \Omega^-\sigma^{0\ast}
+h.c.
\end{align}

\noindent
The neutrino mass matrix in this case has the following form:
\begin{align}
{m_{\nu}}_{ab}\sim \frac{1}{(16\pi^2)^2}
\frac{\lambda\mu\langle \sigma^0\rangle \langle H^0 \rangle}{\Lambda^2}
y^{\eta}_{ai}y^{\chi}_{ij}y^{S}_{jb}. 
\label{mnu:T-II-S-i}
\end{align}

\FloatBarrier
\begin{table}[b!]
\centering
\footnotesize
\resizebox{0.5\textwidth}{!}{
\begin{tabular}{|c|c|}
\hline
Topology &$SU(2)_L\times U(1)_Y\times U(1)_{B-L}$  \\ \hline\hline
T-II-S-iii&
\pbox{10cm}{
\vspace{2pt}
$\sigma^0(1,0,\textcolor{blue}{3})$\\
$S^+(1,1,\textcolor{blue}{5})$\\
$\chi^{++}(1,2,\textcolor{blue}{2})$\\
$\eta(2,\frac{1}{2},\textcolor{blue}{0})$\\
$\zeta(2,-\frac{3}{2},\textcolor{blue}{-2})$
\vspace{2pt}}
\\ \hline
\end{tabular}
}
\caption{ 
Quantum numbers of the BSM scalars.
}\label{tab:T-II-S-iii}
\end{table}
\begin{figure}[b!]
\centering
\includegraphics[scale=0.4]{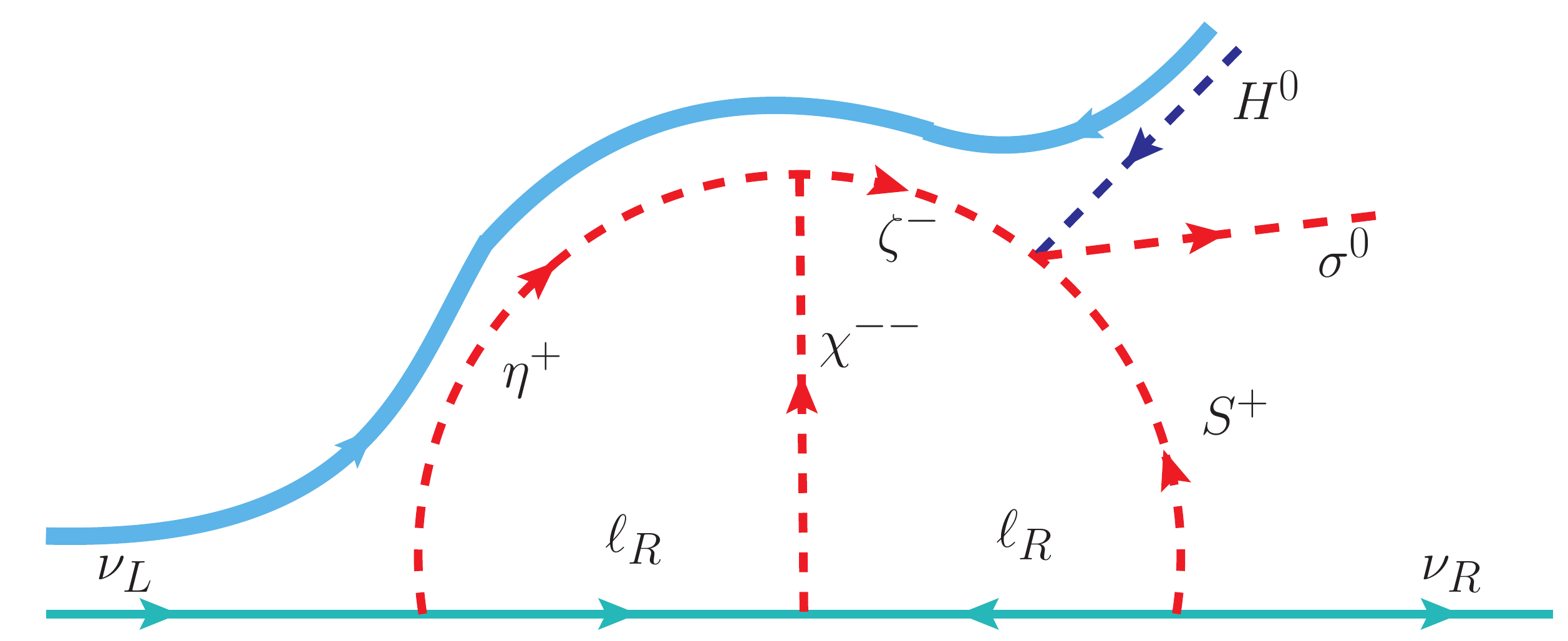}
\caption{Two-loop Dirac neutrino mass for the particle content shown in Table \ref{tab:T-II-S-iii}. }\label{fig:T-II-S-iii}
\end{figure}   

For topology T-II-S-iii, the model is presented in Table. \ref{tab:T-II-S-iii}, and the corresponding Feynman diagram in Fig. \ref{fig:T-II-S-iii}. 
Model with  topology T-II-S-iii can be found by replacing the singly charged scalar $\Omega^+$ of the previous model (Table \ref{tab:T-II-S-i})  by an iso-spin doublet $\zeta=(\zeta^-\;\zeta^{--})^T$ having hypercharge of $-3/2$. This doublet that can not acquire VEV and also cannot couple directly to the fermions, carries non-zero $B-L$ charge.   
 For this model, Yukawa Lagrangian is identical to that of Eq. \eqref{LY:T-II-S-i}.
Then the relevant part of the scalar potential responsible for completing the loop diagram is given by:
\begin{align}
V\supset 
\mu\;\zeta^{\dagger}\eta \chi^{--}
+
\lambda\;H\epsilon \zeta S^+\sigma^{0\ast}
+h.c.
\end{align}

\noindent
In this case, the Dirac neutrino mass matrix also has the same structure as that of  T-II-S-i given in Eq. \eqref{mnu:T-II-S-i}.

\FloatBarrier
\begin{table}[b!]
\centering
\footnotesize
\resizebox{0.5\textwidth}{!}{
\begin{tabular}{|c|c|}
\hline
Topology &$SU(2)_L\times U(1)_Y\times U(1)_{B-L}$  \\ \hline\hline
T-II-S-ii&
\pbox{10cm}{
\vspace{2pt}
$\sigma^0(1,0,\textcolor{blue}{3})$\\
$S^+(1,1,\textcolor{blue}{5})$\\
$\chi^{++}(1,2,\textcolor{blue}{2})$\\
$\eta(2,\frac{1}{2},\textcolor{blue}{0})$\\
$\xi^+(1,1,\textcolor{blue}{2})$
\vspace{2pt}}
\\ \hline
\end{tabular}
}
\caption{ 
Quantum numbers of the BSM scalars.
}\label{tab:T-II-S-ii}
\end{table}
\begin{figure}[b!]
\centering
\includegraphics[scale=0.4]{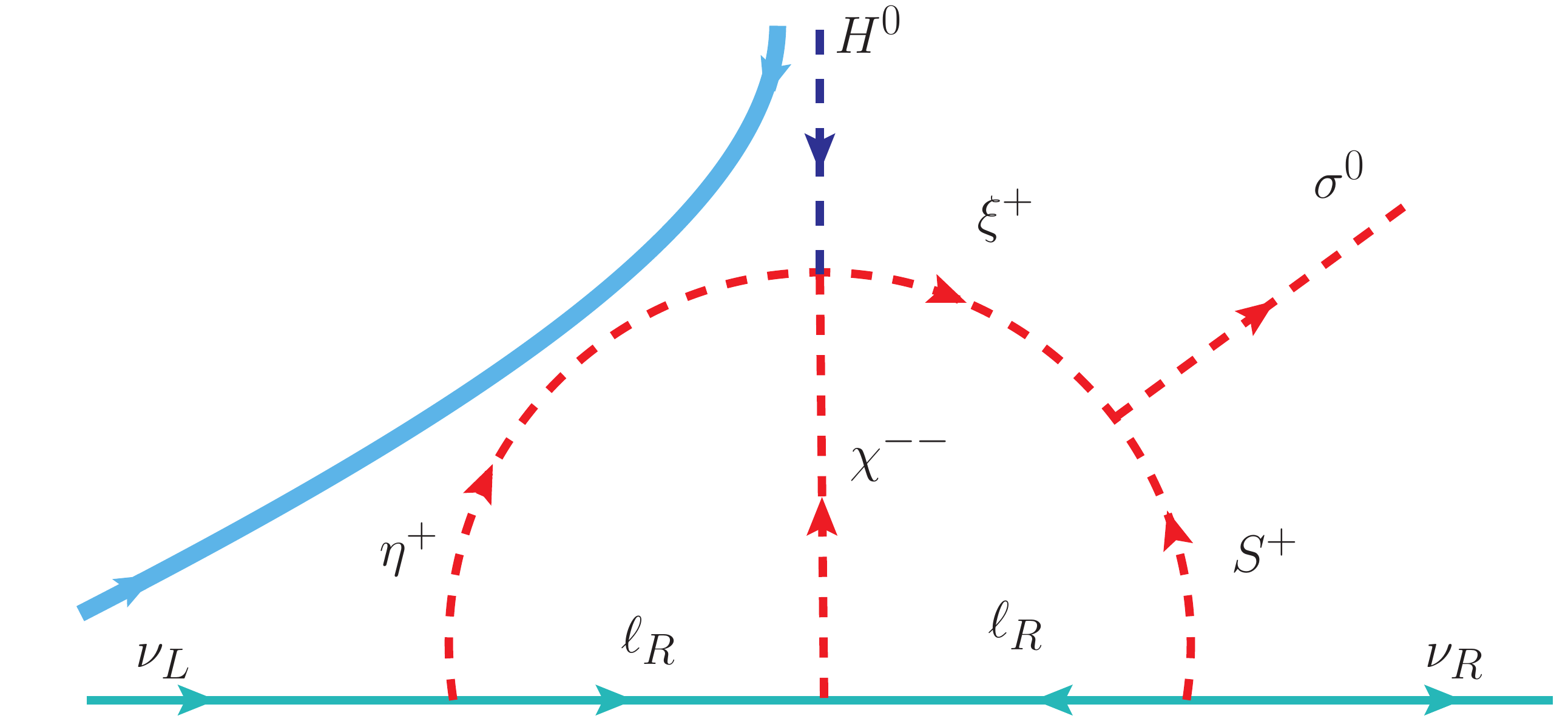}
\caption{Two-loop Dirac neutrino mass for the particle content shown in Table \ref{tab:T-II-S-ii}. }\label{fig:T-II-S-ii}
\end{figure}   

And finally for the  topology of type T-II-S-ii, the needed particle representations are presented in Table. \ref{tab:T-II-S-ii} and the corresponding Feynman diagram is presented in Fig. \ref{fig:T-II-S-ii}.  However, note that for the realization of this topology,  singly charged scalar $\xi^+$ which carries $+2$ unit of $U(1)_{B-L}$ charge is needed to complete the loop. Due to the presence of this multiplet that allows the additional Yukawa coupling $y^{\xi}_{ij}\overline{L_L^c}_i\epsilon{L_L}_j\xi^+$ (not present in the previous two models) on the top of the terms already present  in Eq. \eqref{LY:T-II-S-i},  this model automatically accommodates Dirac neutrino mass at the one-loop level given in Fig. \ref{fig:T-I-F-i}. Hence the T-II-S-ii topology can not lead to a true two-loop model.

From the Yukawa interactions of the two-loop models, Eq. \eqref{LY:T-II-S-i} one sees that compared to the one-loop model, more parameters are involved in the neutrino mass generation.  Even though  the parameter space is somewhat relaxed compared to the one-loop scenario, there exist non-trivial restrictions on the parameters  from experimental bounds on various processes. 
As already pointed out, phenomenologies associated with the Yukawa couplings $y^S_{ij}$  of the singly charged scalar $S^+$ are studied in \cite{Nasri:2001ax, Kanemura:2011jj}. The Yukawa couplings $y^{\chi}$ of the doubly charged scalar $\chi^{++}$ with the fermions  contribute to interesting flavor violating processes as well. Furthermore, 
due to the presence of the second Higgs doublet, the off-diagonal couplings of the iso-doublets with the SM fermions  cannot be rotated away, hence lead to  LVF processes. Few of the most interesting LFV processes are:  radiative rare decays $\ell_a\to \ell_b \gamma$,  tree-level tri-lepton decays $\ell_a\to \ell_b\overline{\ell}_c\ell_d$ and $\mu - e$ conversion in the nuclei.  Some of these processes put stringent constraints on the off-diagonal Yukawa couplings. Whereas, flavor conserving processes give rise to electron and muon electric dipole moment and also anomalous magnetic moments via loop diagrams.  The phenomenology of this class of models are very similar to the Zee model and Zee-Babu model and  detail phenomenological studies of these processes including LHC signals have been carried out in many different works. Instead of listing all, we refer the reader to the most recent analysis   \cite{Herrero-Garcia:2014hfa, Herrero-Garcia:2017xdu} and the references therein.

Here, for a demonstration we  only explore the most constraining process which is $\mu\to e\gamma$ rare decay.
It is beyond the scope of this paper to explore the full parameter space and to incorporate all the existing bounds simultaneously, which is also not the purpose of this present paper.   In this analysis,  for simplicity we assume that the Yukawa couplings of the doubly charged scalar is somewhat smaller than the couplings with the iso-spin doublets and furthermore, its mass $m_{\chi}$ is of the order of several TeV or higher. Consequently its contribution to $\mu\to e\gamma$ is neglected in the following. With these assumptions $\mu\to e\gamma$ puts bounds on the Yukawa couplings and the masses of the scalars associated with the iso-doublets that we assume to live in the sub-TeV range.

One of the singly charged scalars from the two Higgs doublets will be eaten up by the corresponding gauge boson, we denote the remaining physical state by $h^+$. We also denote the two CP-even  states as $h_{1,2}$ ($h_1$ is identified as the SM Higgs of mass 125 GeV) and the only CP-odd state by $A$.  We assume the doublets $H$ and $\eta$ take VEVs $v_1$ and $v_2$ respectively (with  $v^2_1+v^2_2=246^2$ GeV$^2$) and define $t_{\beta}\equiv \tan \beta = v_2/v_1$.  Then going to a basis where the charged lepton mass matrix is diagonal, the relevant Lagrangian takes the following form: 
\begin{align}
-\mathcal{L}_Y &\supset
y^{h^+}\overline{\nu_L}e_Rh^+
+i\;y^{A}\overline{e_L}e_RA
+y^{h_1}\overline{e_L}e_Rh_1
+y^{h_2}\overline{e_L}e_Rh_2 +h.c.
\end{align}

\noindent
With,
\begin{align}
&y^{h_2}=  \left( \frac{m_Ec_{\alpha}}{vc_{\beta}}-s_{\beta-\alpha}\frac{y^{\eta}}{\sqrt{2}c_{\beta}} \right),\;\;\;
y^{h_1}= \left( \frac{-m_Es_{\alpha}}{vc_{\beta}}+c_{\beta-\alpha}\frac{y^{\eta}}{\sqrt{2}c_{\beta}} \right),\\
&y^{A}=  \left( -\frac{m_Et_{\beta}}{v}+\frac{y^{\eta}}{\sqrt{2}c_{\beta}} \right),\;\;\;
y^{h^+}=  \left( \frac{-\sqrt{2}m_Et_{\beta}}{v}+\frac{y^{\eta}}{c_{\beta}} \right). 
\end{align}

\noindent Here $\alpha$ represents the mixing angle associated with the two CP-even states. 
From the effective Lagrangian of the form,
\begin{align}
\mathcal{L}_{eff}=\frac{e}{2}\overline{\psi} \left( C^LP_L+C^RP_R  \right)\sigma_{\mu\nu}\psi F_{\mu\nu},
\end{align}

\noindent
the matrix element for the decay $\ell_i\to\ell_j\gamma$ can be written as:
\begin{align}
M(\ell_i\to\ell_j\gamma)=ie\overline{\psi}_j\left(  C_{ji}^LP_L+C_{ji}^RP_R \right)
\sigma_{\mu\nu}\psi_i\epsilon^{\ast}_{\mu}q_{\nu}.
\end{align}

\noindent
Then the decay width for the process $\ell_i\to\ell_j\gamma$ is found to be \cite{Lavoura:2003xp}:
\begin{align}
\Gamma(\ell_i\to\ell_j\gamma)=\frac{\alpha_{em}m^3_{\ell_i}}{4}
\left( |C^L_{ji}|^2 + |C^R_{ji}|^2 \right).
\end{align}

\noindent
From the Lagrangian given above, the $C^{L,R}$ coefficients can be computed as:   
\begin{align}
C^L=-\frac{1}{16\pi^2}\left( \frac{c^L_{h_2}}{m^2_{h_2}}+ \frac{c^L_{h_1}}{m^2_{h_1}}+ \frac{c^L_A}{m^2_A}+ \frac{c^L_{h^+}}{m^2_{h^+}}  \right),\;\;\; C^R=\left( C^L  \right)^{\dagger},
\end{align}

\noindent where we find,
\begin{align}
&c^L_{\phi^0}=
y^{\phi^0}F^{\phi^0}_1(y^{\phi^0})^{\dagger}m_E
+m_E(y^{\phi^0})^{\dagger}F^{\phi^0}_1y^{\phi^0}
+(y^{\phi^0})^{\dagger}m_EF^{\phi^0}_3(y^{\phi^0})^{\dagger},\;\;\;\phi^0=h_2,h_1,A.
\\
&c^L_{\phi^+}=
(y^{\phi^+})^{\dagger}F^{\phi^+}_2y^{\phi^+}m_E,\;\;\; \phi^+=h^+.
\end{align}

\FloatBarrier
\begin{figure}[t!]
\begin{center}
\includegraphics[width=7.0cm]{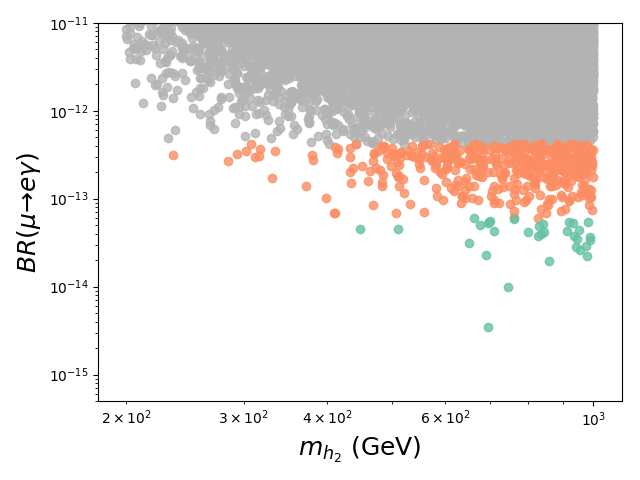} 
\includegraphics[width=7.0cm]{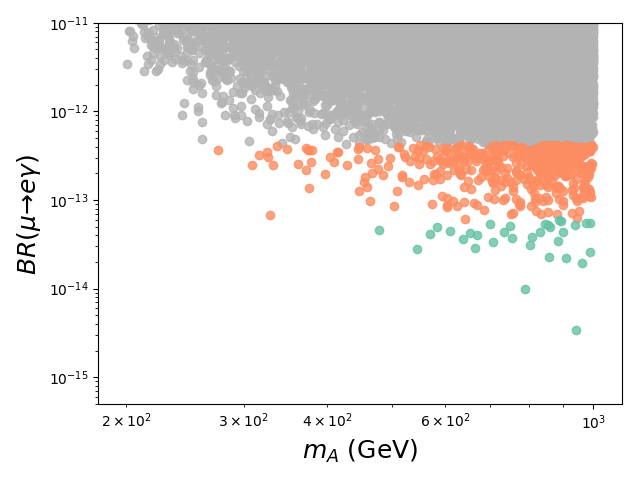} \\
\includegraphics[width=7.0cm]{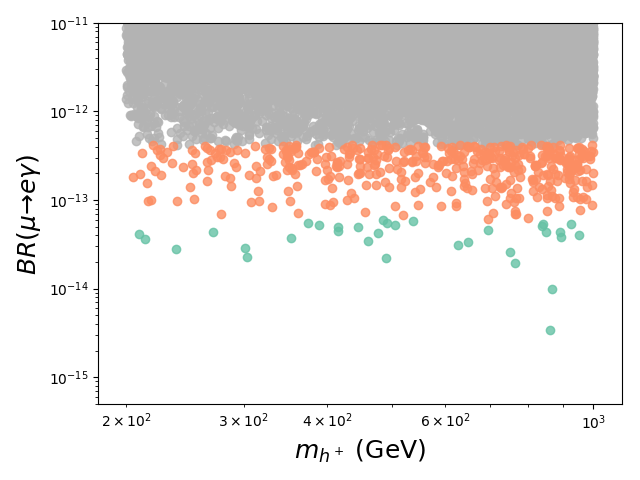} 
\end{center}
\caption{In these plots the Yukawa couplings $|y^{\chi}_{22,23,32}|$ are randomly varied within the range $10^{-4}-10^{-3}$, $|y^{\chi}_{11,12,21,13,31}|$ are  within range $10^{-6}-10^{-4}$ and  $BR(\mu \to e\gamma)$ is plotted as a  function of the  masses of the associated scalars. Gray dots belong to the region  that is experimentally ruled out, the current experimental upper bound is $BR(\mu \to e\gamma)< 4.2\times 10^{-13}$ (MEG collaboration)  and the projected sensitivity is $BR(\mu \to e\gamma)\leq 6.0\times 10^{-14}$ (MEG II collaboration) \cite{TheMEG:2016wtm}. Within the experimentally allowed region,   
 dots with orange color [green] correspond to $ 6.0 \times 10^{-14} <BR(\mu \to e\gamma)\leq 4.2\times 10^{-13}$ [$BR(\mu \to e\gamma)\leq 6.0\times 10^{-14}$]. }
\label{BR-II}
\end{figure}

\noindent
And the loop functions are given by \cite{Lavoura:2003xp}: 
\begin{align}
&F^{\phi}_k=diag \left[
F_k\left( \frac{m^2_{e}}{m^2_{\phi}}\right), 
F_k\left( \frac{m^2_{\mu}}{m^2_{\phi}}\right), 
F_k\left( \frac{m^2_{\tau}}{m^2_{\phi}}\right) 
\right],
\;\;\; \phi= \phi^0, \phi^+,
\\
&F_1(z)=\frac{z^2-5z-2}{12(z-1)^3}+\frac{z\text{ln}z}{2(z-1)^4},\\
&F_2(z)=\frac{2z^2+5z-1}{12(z-1)^3}-\frac{z^2\text{ln}z}{2(z-1)^4},\\
&F_3(z)=\frac{z-3}{2(z-1)^2}+\frac{\text{ln}z}{(z-1)^3}.
\end{align}

In the basis we are working where the charged lepton mass matrix is diagonal, the Yukawa coupling matrix $y^{\eta}$ is a general complex matrix. For computing the branching ratio $Br(\mu\to e\gamma)$, we randomly vary the entries of this matrix within the range: $|y^{\eta}_{ij}|=10^{-6}-10^{-4}$ (for $ij=11,12,13,21,31$) and  $|y^{\eta}_{ij}|=10^{-4}-10^{-3}$ (for $ij=22,23,32$). The corresponding scatter plot is presented in Fig. \ref{BR-II} as a function of the physical scalars that emerge from the iso-spin doublets and participate in $\mu\to e\gamma$ process.

\section{Minimal three-loop models}\label{three}
Following the same spirit of the one-loop and two-loop models, one can also build models for three-loop Dirac neutrino mass within our framework. 
In this section we do a systematic analysis for the three-loop scenario and 
show that minimal three-loop model can be constructed 
with just five BSM scalars, this number is the same as that of the  two-loop models presented in the previous section. 

\begin{figure}[t!]
\centering\includegraphics[scale=0.35]{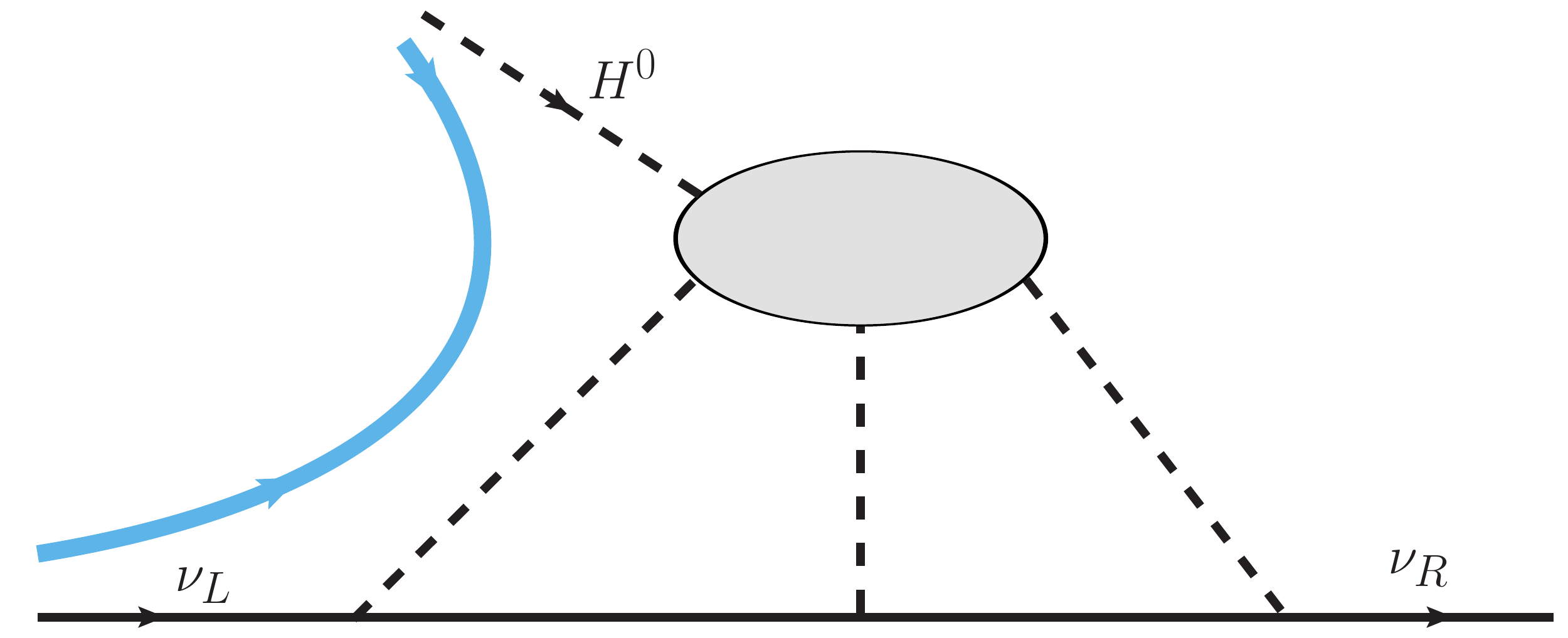}
\caption{ Expected three-loop topology of type T-III-S-x within our framework.  The gray shaded region contains a loop in it that dpends on the details of the model. }\label{three-loop-B}
\end{figure}

For a systematic study of the three-loop scenario   in search of the minimal models, in coherence with the one-loop and two-loop cases, we proceed to construct models utilizing the same  dimension-5 operator given in Eq. \eqref{operator}. Following the discussion of two-loop models, it is straightforward to find the expected generic  topology for the three-loop models which is presented in Fig.  \ref{three-loop-B} that is of the type T-III-S. The other possibility of attaching the SM Higgs doublet with the fermion propagator cannot give rise to a true three-loop model in our  set-up as already pointed out in the previous section, hence we discard  such possibility here.

\FloatBarrier

\begin{figure}[b!]
\centering\includegraphics[scale=0.35]{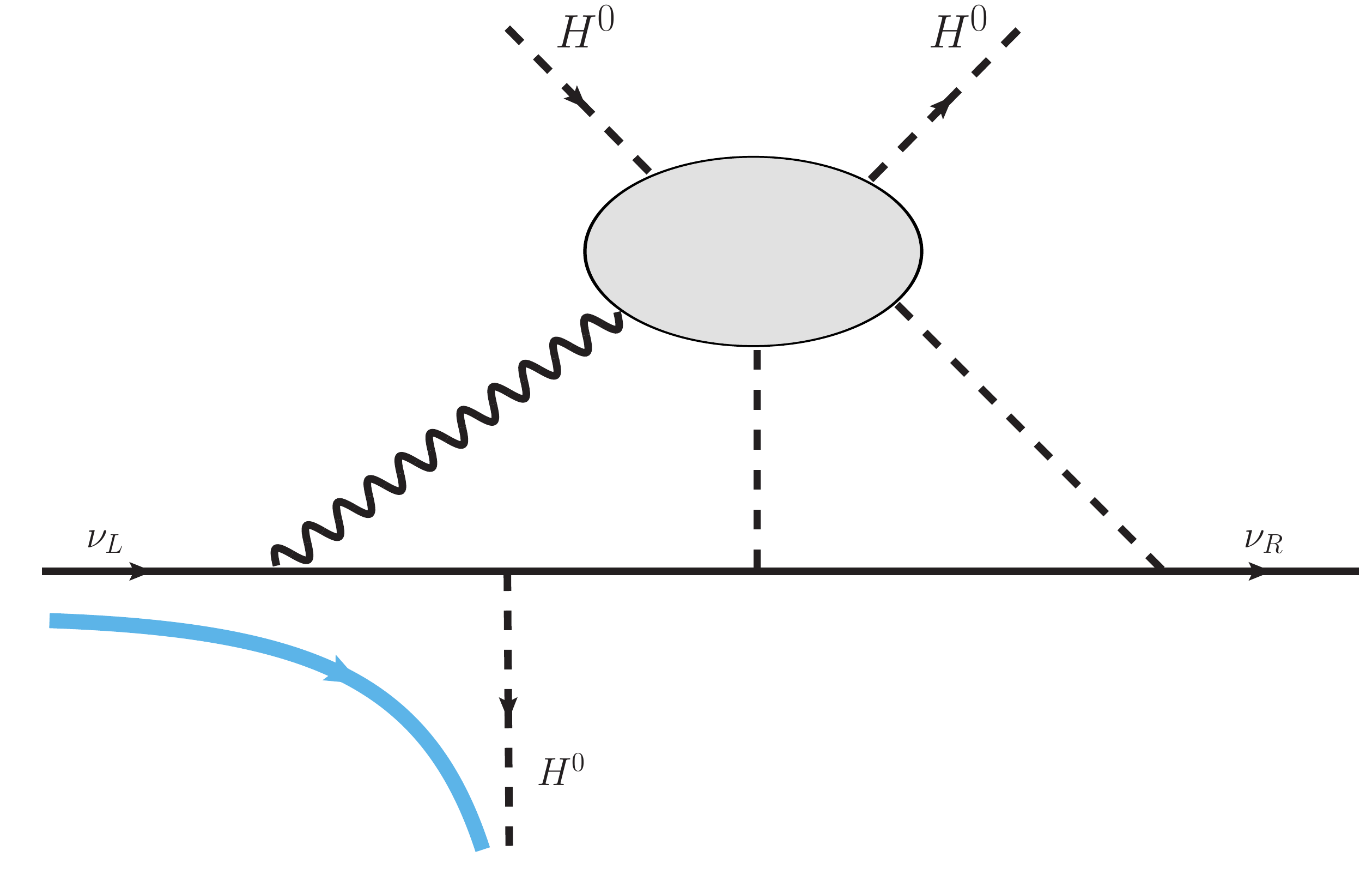}
\caption{ Three-loop topologies T-III-F-x within our framework. The direction of the flow of the iso-spin doublet is unique in this scenario. The gray shaded  region   contains a loop in it. }\label{three-loop}
\end{figure}

However, we will show  that three-loop models having the topology T-III-S-x are not economical and require at least seven BSM scalars. On the other hand, minimal three-loop models can be built by considering a different topology that makes use of a  vertex involving gauge boson - fermion interaction already present in the SM. Utilization of such a vertex allows us to construct 
 minimal model that requires only five BSM scalars instead of seven.  This topology is unique within our set-up and  exhibited in Fig. \ref{three-loop} which we label  as T-III-F. However, constructing diagrams with this topology requires the implementation of the dimension-7 operator  introduced  in Eq. \eqref{operator-7} instead of dimension-5.  We first present the minimal three-loop model below by making  use of this dimension-7 operator  and  later discussion the non-minimality of models associated with dimension-5 operator  by explicitly constructing  three-loop models in Appendix \ref{non-min}.

\FloatBarrier

\begin{figure}[th!]
\centering
\includegraphics[scale=0.4]{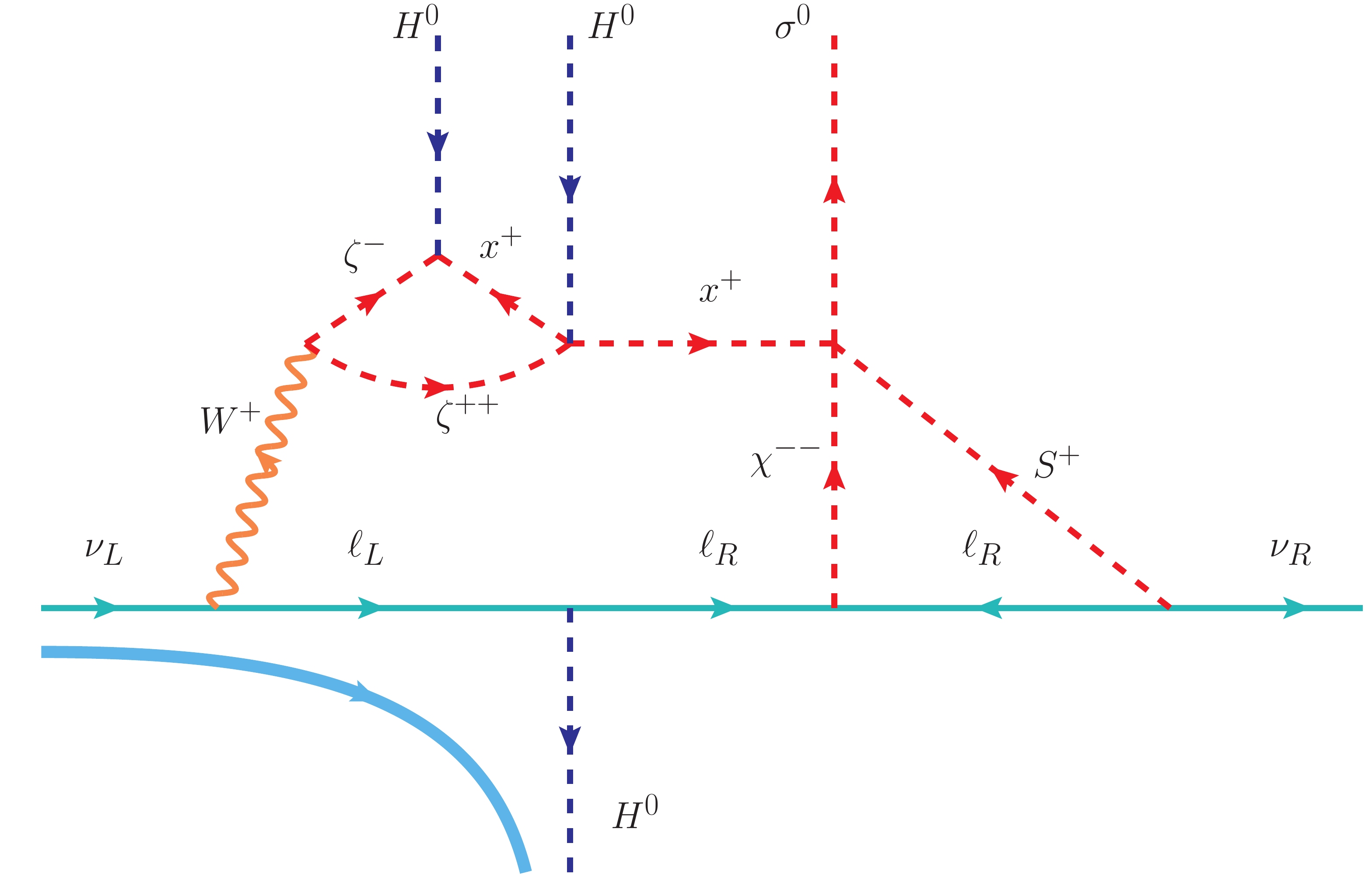}
\caption{Three-loop Dirac neutrino mass for the particle content shown in Table \ref{tab:T-III-F-i}. }\label{fig:T-III-F-i}
\end{figure}

Looking at the topology presented in Fig. \ref{three-loop}, it is not difficult to understand that the direction of the  flow of the iso-spin doublet is unique as a result of making use of the SM gauge boson interactions with the SM fermions.  Utilizing the associated dimension-7 operator in our set-up,    the minimal model for three-loop realization having the topology T-III-F-i is presented in Table \ref{tab:T-III-F-i} with the corresponding Feynman diagram shown in Fig. \ref{fig:T-III-F-i}.  Completion of this diagram requires two singly charged scalars, one of them with zero $B-L$. Furthermore,   a doubly charged scalar and an iso-doublet with hypercharge $-3/2$ neutral under $B-L$. Unlike the two-loop models, a second SM like Higgs doublet is not required here.

\FloatBarrier
\begin{table}[th!]
\centering
\footnotesize
\resizebox{0.5\textwidth}{!}{
\begin{tabular}{|c|c|}
\hline
Topology &$SU(2)_L\times U(1)_Y\times U(1)_{B-L}$  \\ \hline\hline
T-III-F-i&
\pbox{10cm}{
\vspace{2pt}
$\sigma^0(1,0,\textcolor{blue}{3})$\\
$S^+(1,1,\textcolor{blue}{5})$\\
$\chi^{++}(1,2,\textcolor{blue}{2})$\\
$\zeta(2,-\frac{3}{2},\textcolor{blue}{0})$\\
$x^+(1,1,\textcolor{blue}{0})$
\vspace{2pt}}
\\ \hline
\end{tabular}
}
\caption{ 
Quantum numbers of the BSM scalars.
}\label{tab:T-III-F-i}
\end{table}
\FloatBarrier

\begin{figure}[t!]
\centering
\includegraphics[scale=0.4]{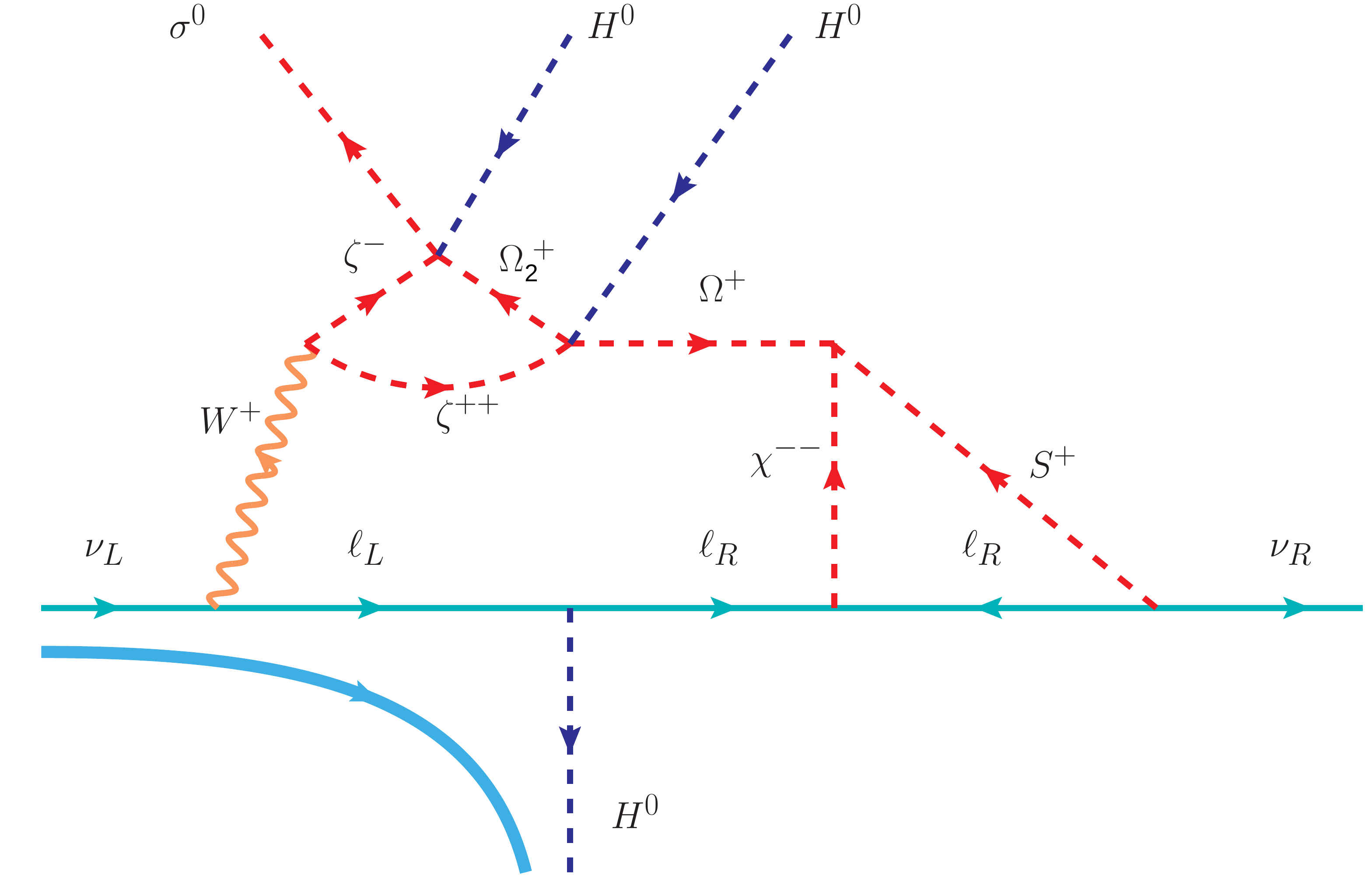}
\caption{Three-loop Dirac neutrino mass for the particle content shown in Table \ref{tab:T-III-F-ii}.  
By a trivial $B-L$ charge swapping: $\zeta(2,-\frac{3}{2},\textcolor{blue}{0})\to$ $\zeta(2,-\frac{3}{2},\textcolor{blue}{3})$ and $\Omega^{+}_2(1,1,\textcolor{blue}{3})\to$  $\Omega^{+}_2(1,1,\textcolor{blue}{0})$  another model (with topology T-III-F-ii-b) can be constructed. }\label{fig:T-III-F-ii}
\end{figure}

We also present two next-to-minimal models that share the same diagram (corresponding topologies are denoted by T-III-F-ii-a and T-III-F-ii-b) but with different set of scalar  multiplets. This second set of topologies are variant of the minimal three-loop scenario (T-III-F-i) and each requires six BSM scalar multiplets. The particle content of these models are presented in   Table \ref{tab:T-III-F-ii}  and the associated Feynman diagram is presented  in Fig. \ref{fig:T-III-F-ii}.

\FloatBarrier
\begin{table}[t!]
\centering
\footnotesize
\resizebox{0.5\textwidth}{!}{
\begin{tabular}{|c|c|}
\hline
Topology &$SU(2)_L\times U(1)_Y\times U(1)_{B-L}$  \\ \hline\hline
T-III-F-ii-a&
\pbox{10cm}{
\vspace{2pt}
$\sigma^0(1,0,\textcolor{blue}{3})$\\
$S^+(1,1,\textcolor{blue}{5})$\\
$\chi^{++}(1,2,\textcolor{blue}{2})$\\
$\zeta(2,-\frac{3}{2},\textcolor{blue}{0})$\\
$\Omega^+(1,1,\textcolor{blue}{-3})$\\
$\Omega^{+}_2(1,1,\textcolor{blue}{3})$
\vspace{2pt}}
\\ \hline
\end{tabular}
}
\caption{ 
Quantum numbers of the BSM scalars. A simple variation of this model can be  constructed  with same topology (denoted by T-III-F-ii-b) by replacing 
$\zeta(2,-\frac{3}{2},\textcolor{blue}{0})\to$ $\zeta(2,-\frac{3}{2},\textcolor{blue}{3})$ and $\Omega^{+}_2(1,1,\textcolor{blue}{3})\to$  $\Omega^{+}_2(1,1,\textcolor{blue}{0})$. 
}\label{tab:T-III-F-ii}
\end{table}
\FloatBarrier
 
All these three-loop models share the same Yukawa sector:
\begin{align}
\mathcal{L}_Y \supset
y^{H}_{ij}\overline{L_L}_i{\ell_R}_jH
+
y^{S}_{kj}\overline{\nu_R^c}_k{\ell_R}_jS^+
+
y^{\chi}_{ij}\overline{\ell_R^c}_i{\ell_R}_j\chi^{++}
+h.c.
\label{LY:T-III-F}
\end{align}

\noindent
The relevant part of the scalar potential for the model with topology T-III-F-i is:
\begin{align}\label{T-III}
V\supset 
\mu \zeta \epsilon H x^+ + 
\lambda \chi^{--}x^+S^+\sigma^{0\ast}
+ \lambda^{\prime} \zeta^{\dagger}H x^-x^-
+ h.c.
\end{align}
\noindent
whereas for topology of type T-III-F-ii-a(b) it is:
\begin{align}
V\supset 
\lambda \zeta \epsilon H \Omega^{+}_2\sigma^{0\ast}
 + 
\lambda^{\prime}\zeta^{\dagger}H\Omega^-\Omega^{-}_2
+\mu\chi^{--} S^+ \Omega^+
+ h.c.
\end{align}
\noindent
All these three-loop models have the same form for the neutrino mass matrix given by: 
\begin{align}
{m_{\nu}}_{ab}\sim
\frac{1}{(16\pi^2)^3} \frac{g^2\lambda \lambda^{\prime} \mu \langle \sigma^0 \rangle \langle H^0 \rangle^3}{\Lambda^4} y^H_{ai}  y^{\chi}_{ij}  y^S_{jb}.
\end{align}

\noindent
Here $g$ represents the $SU(2)$ gauge coupling constant. Two of the neutrinos receive non-zero mass after the $U(1)_{B-L}$ and the EW symmetries are broken. The reason for the masslessness of the two chiral states associated with $\nu_{R_1}$ is the same as that of  the two-loop case.

\FloatBarrier
\begin{figure}[b!]
\begin{center}
\includegraphics[width=7.5cm]{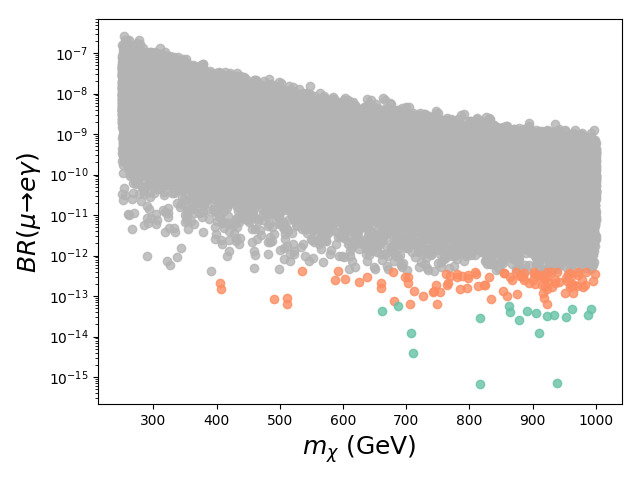} \hspace{0.1in}
\includegraphics[width=7.5cm]{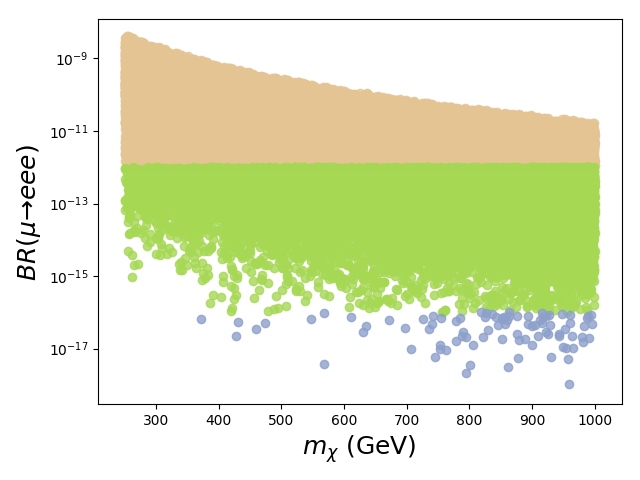}
\end{center}
\caption{In these plots we randomly vary $|y^{\chi}_{11,12,13}|$  within the range $10^{-4}-10^{-2}$ and  $|y^{\chi}_{22,23}|$  within  $0.1-0.9$. These Yukawa couplings are marginalized while plotting the branching ratios as a function of the mass of the doubly charged scalar that is varied in between 250 GeV to 1 TeV.  For the process $\mu \to eee$, the present bound is $10^{-12}< BR(\mu \to eee)$ \cite{Bellgardt:1987du} and future sensitivity is $10^{-16}< BR(\mu \to eee)$ \cite{Blondel:2013ia}.  In the left plot, dots with orange color [green] correspond to $6.0 \times 10^{-14} <BR(\mu \to e\gamma)\leq 4.2\times 10^{-13}$ [$BR(\mu \to e\gamma)\leq 6.0\times 10^{-14}$].
In the right plot, dots with light green color [blue] correspond to $ 1.0 \times 10^{-12} <BR(\mu \to eee)\leq 1.0\times 10^{-16}$ [$BR(\mu \to eee)\leq 1.0\times 10^{-16}$].  }
\label{BR-III}
\end{figure}

The phenomenology of the minimal three-loop model  has some similarity with the one-loop and two-loop models. Among  the Yukawa interactions given in Eq. \eqref{T-III}, $y^S$ and $y^{\chi}$ can in principle mediate LFV processes. Like all the other minimal models of this work, $y^S_{i1}=0$ due to different $U(1)_{B-L}$ charge assignment of ${\nu_R}_1$, hence the Yukawa coupling $y^S$ is phenomenologically less interesting.  However,  the Yukawa interactions associated with the doubly charged scalar $y^{\chi}$ can lead to dangerous LFV. Here we impose the most stringent experimental bounds on the relevant parameter space which is coming from the $\mu\to e\gamma$ branching ratio. We also simultaneously impose bounds from the rare three lepton final states, among  different such processes $\mu\to eee$ is the most constraining one.       Recall that  $y^{\chi}_{ij}$  is a symmetric matrix, and here for simplicity we ignore any mixing of the $\chi$ multiplet with the other doubly charged states.  The associated branching ratios in the present model are given by: 
\begin{align}
&Br(\mu\to e\gamma)=\frac{\alpha_{em}}{3\pi G^2_F m^4_{\chi}}
\left| (y^{\chi\dagger}y^{\chi})_{\mu e}  \right|^2,  \\
&Br(\mu\to eee)=\frac{1}{4G^2_F m^4_{\chi}}
\left| y^{\chi}_{\mu e}y^{\chi\ast}_{ee}  \right|^2. 
\end{align}

By randomly varying the Yukawa couplings $|y^{\chi}_{22,23}|$ within the range $0.1-0.9$ and $|y^{\chi}_{11,12,13}|$ within the range $10^{-4}-10^{-2}$, we plot these branching ratios as a function of the mass of the doubly charged scalar in  Fig. \ref{BR-III}.

\vspace{20pt}
\noindent
\textbf{Dark Matter candidate in this class of models}\\
Here we briefly discuss the possibility of a DM candidate  in the class of models presented in this work. Even though we do not demand the presence of a DM in our set-up, one can readily get a DM candidate if a Majorana mass term is allowed only for the ${\nu_R}_1$ state that carries a different charge under $U(1)_{B-L}$. As discussed in great details, in this class of models, the chiral states associated with ${\nu_R}_1$ cannot get a Dirac mass if the  minimal models are not extended by more scalar multiplets (such extensions will be discussed in appendices \ref{A}, \ref{B} and \ref{C} for one-loop, two-loop and three-loop models).   From the structures of the bilinears listed  in Sec. \ref{SEC-02}, one can see that introducing a multiplet with quantum number $\sigma^{0}_4(1,0,+10)$ can allow a Majorana mass of the form: $\overline{{\nu_R^c}_1}{\nu_R}_1 \sigma^{0}_4$. Once this scalar gets a VEV, ${\nu_R}_1$ receives a Majorana mass and can serve as a Majorana DM candidate \cite{Nomura:2017jxb}. In such a scenario, the imaginary part of $\sigma^{0}_4$ will become a physical Goldstone bosons due to an accidental global $U(1)$ symmetry respected by the scalar potential, hence may contribute to $N_{eff}$. However, this does not cause any serious problem since, this scalar does not couple directly to the SM fermions except the SM Higgs doublet. It can decouple from the thermal bath early in the universe if its coupling with the SM Higgs is somewhat suppressed $\lesssim 10^{-3}$, for details see for example Refs.  \cite{Weinberg:2013kea, Davidson:2009ha, Baek:2016wml, Nomura:2017jxb}.

Finally, we compare the simplest models presented here with the existing models of radiative Dirac neutrino mass in the literature that use only $U(1)_{B-L}$ symmetry.  In  \cite{Wang:2017mcy, Calle:2018ovc, Bonilla:2018ynb}, the one-loop realization requires two BSM chiral fermions and three scalar multiplets. Similarly, two-loop realizations require more BSM states compared to our scenario here, in \cite{Wang:2017mcy} four chiral fermions and four scalars are introduced for two-loop implementation. On the other hand, three-loop model is not realization in the literature before.   We remind the readers that within our framework, implementation of minimal one-loop model requires  three, and minimal  two-loop and three-loop models require only five BSM scalar multiplets. Hence, the class of models presented in this work are simple and very economical.

\section{Conclusions}\label{SEC-08}
The baryon number minus the lepton number is a global symmetry of the SM that can be made anomaly free by introducing only three right-handed neutrinos $\nu_R$. Due to the presence of the right-handed partners $\nu_R$  of the SM left-handed neutrinos $\nu_L$, it is natural to expect that like rest of the fermions of the SM, neutrinos are also Dirac in nature.
Building models of radiative Dirac neutrino mass usually requires the presence of ad hoc  symmetries  as well as additional fermions beyond the SM. These two requirements introduce extra complexities into the theory. By embracing the point of view that the imposition of ad hoc symmetries is least desirable, and well motivated UV completion of the SM may not contain additional fermions,   
  in this work, we have proposed  new simple  models of radiative Dirac neutrino mass. 
Our framework employs only $U(1)_{B-L}$ symmetry to forbid all the 
  tree-level Dirac and Majorana mass terms for the neutrinos  and does not introduce additional fermions into the theory. 
Within this set-up, we construct minimal models of  Dirac neutrino mass at the 
(i) one-loop, (ii) two-loop and (iii) three-loop level. It is shown that the minimal one-loop model requires three scalar multiplets beyond the SM, whereas minimal two-loop and three-loop models require five.  
The presented two-loop and three-loop models of Dirac neutrino mass have not appeared in the literature before. Possible dark matter candidate in this class of models is briefly discussed.

\section*{Acknowledgments}
The author would like to thank Dr. Ernest Ma and  Vishnu P. K for discussion.

\begin{appendices}
\appendixpageoff
\section{\large Effects of higher dimensional operators: One-loop model}\label{A}
One of the neutrinos remains massless in the class of models discussed in this work. First we stress the fact that one massless neutrino is completely consistent with experimental data. However,  
one may search for the allowed higher dimensional operators  within a given model that may induce mass to the massless chiral pair associated with $\nu_{R_1}$. In this appendix, we discuss the effects of such higher dimensional operators for the minimal one-loop model introduced in Sec. \ref{SEC-01}.    The lowest dimensional operator that can be written down to complete the loop diagram corresponding to the state ${\nu_{R}}_1$ is dimension-6 and is given by:
\begin{align}
\mathcal{L}_6= \frac{h_{i1}}{\Lambda^2}\;\overline{L_L}_i \widetilde{H}\;{\nu_R}_1\left(\sigma^{0\ast}\right)^2. \label{operator-6}
\end{align}

\FloatBarrier
\begin{figure}[th!]
\centering
\includegraphics[scale=0.37]{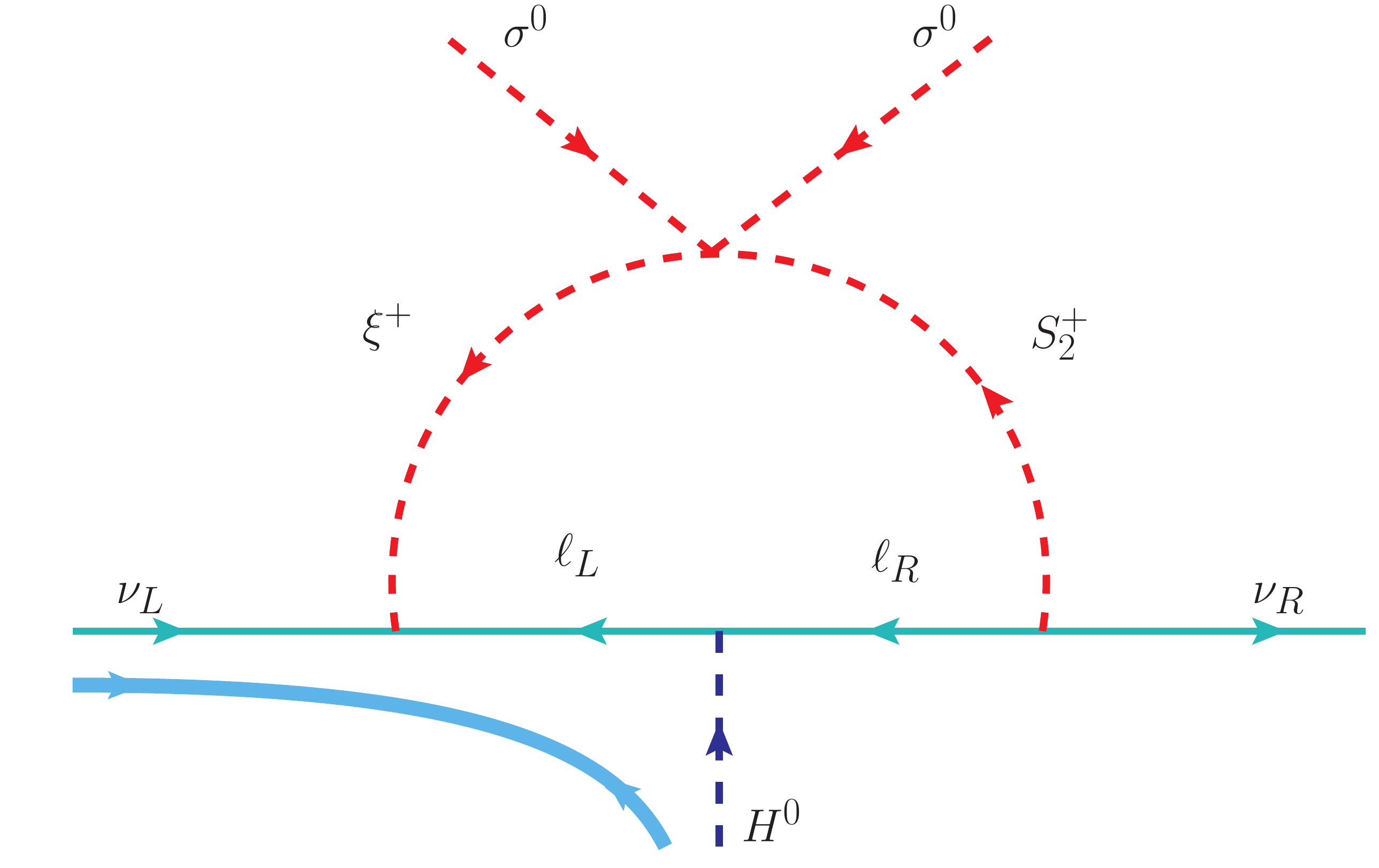}
\caption{One-loop Feynman diagram associated with the state $\nu_{R_1}$ that carries $\textcolor{blue}{+5}$ unit of charge under the $B-L$. This diagram represents the UV completion of the dimension-6 operator given in Eq. \eqref{operator-6}. Completion of this diagram requires one extra singly charged scalar $S^+_2(1,1,\textcolor{blue}{-4})$ in addition to the BSM states listed in Table \ref{tab:T-I-F-i}. }\label{dim6-1}
\end{figure} 

\noindent
The UV complete model of this dimension-6 operator must include an extra singly charged scalar $S^+_2$ with quantum number $(1,1,\textcolor{blue}{-4})$. The corresponding Feynman diagram is presented in Fig. \ref{dim6-1}. It is important to note that in this one-loop model
 no higher dimensional operator can give rise mass to all the neutrinos. Even with the inclusion of this dimension-6 operator (or any other higher dimensional operator), one pair of the chiral states will remain massless which is a direct consequence of the antisymmetric nature of the neutrino mass matrix  given in Eq. \eqref{numass-1}.

\section{\large Effects of higher dimensional operators: Two-loop models}\label{B}
\begin{figure}[th!]
\centering
\includegraphics[scale=0.4]{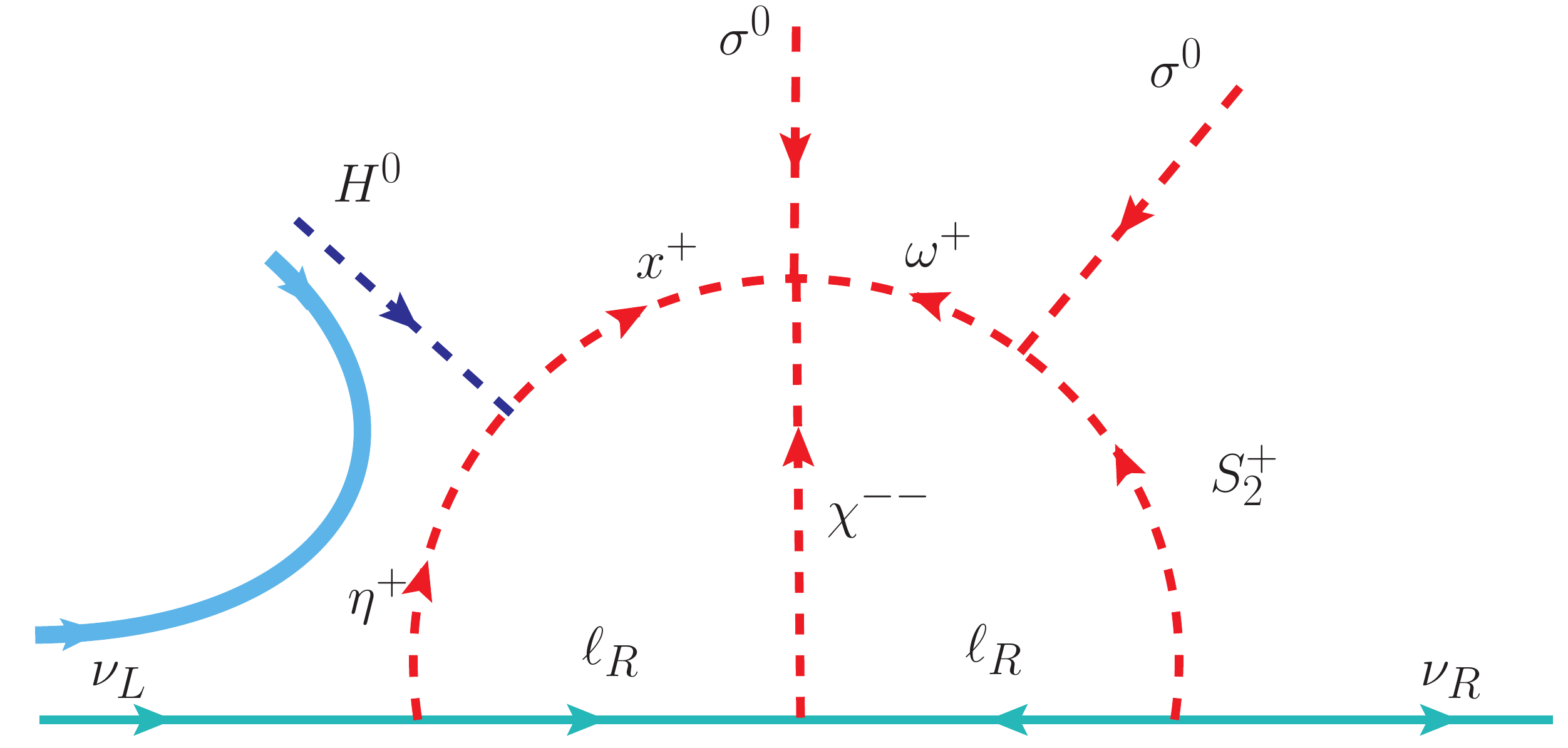}
\caption{Two-loop Feynman diagram associated with the state $\nu_{R_1}$ that carries $\textcolor{blue}{+5}$ unit of charge under the $B-L$ for model with topology T-II-S-i. This diagram represents the UV completion of the dimension-6 operator given in Eq. \eqref{operator-6}. Completion of this diagram requires three extra singly charged scalars $S^+_2(1,1,\textcolor{blue}{-4})$, $\omega^+(1,1,\textcolor{blue}{-1})$ and $x^+(1,1,\textcolor{blue}{0})$  in addition to the BSM states listed in Table \ref{tab:T-II-S-i}. }\label{dim6-2}
\end{figure}

In this appendix, we discuss the relevant higher dimensional operators for the two-loop models  and their UV completions.  Unlike the one-loop model, higher dimensional operators can induce  mass to all the states for these two-loop models. This is a  consequence of the neutrino mass matrix Eq. \eqref{mnu:T-II-S-i}
 not being antisymmetric in nature.  The lowest order operator that can do so   is the dimension-6 operator already introduced  in Eq. \eqref{operator-6}. Let us first consider the two-loop model with topology T-II-S-i defined in Table \ref{tab:T-II-S-i}. The simplest UV completion of this dimension-6 operator requires three additional particles on top of the multiplets listed in Table \ref{tab:T-II-S-i}. It requires three singly charged particles: $S^+_2(1,1,\textcolor{blue}{-4})$, $\omega^+(1,1,\textcolor{blue}{-1})$ and $x^+(1,1,\textcolor{blue}{0})$. The corresponding Feynman diagram is presented in Fig. \ref{dim6-2}.  

\FloatBarrier

Alternatively, one can search for a more economical way to induce this mass. However, this requires the presence of a second SM singlet (other than $\sigma^0$) into the theory.  This will allow a  new dimension-6 operator not present within the minimal set-up and is  given by:
\begin{align}
\mathcal{L}_6= \frac{h_{i1}}{\Lambda^2}\;\overline{L_L}_i \widetilde{H}\;{\nu_R}_1  \left(\sigma^{0}   \sigma^{0\ast}_2\right). \label{operator-6-B}
\end{align}  

\FloatBarrier
\begin{figure}[b!]
\centering
\includegraphics[scale=0.4]{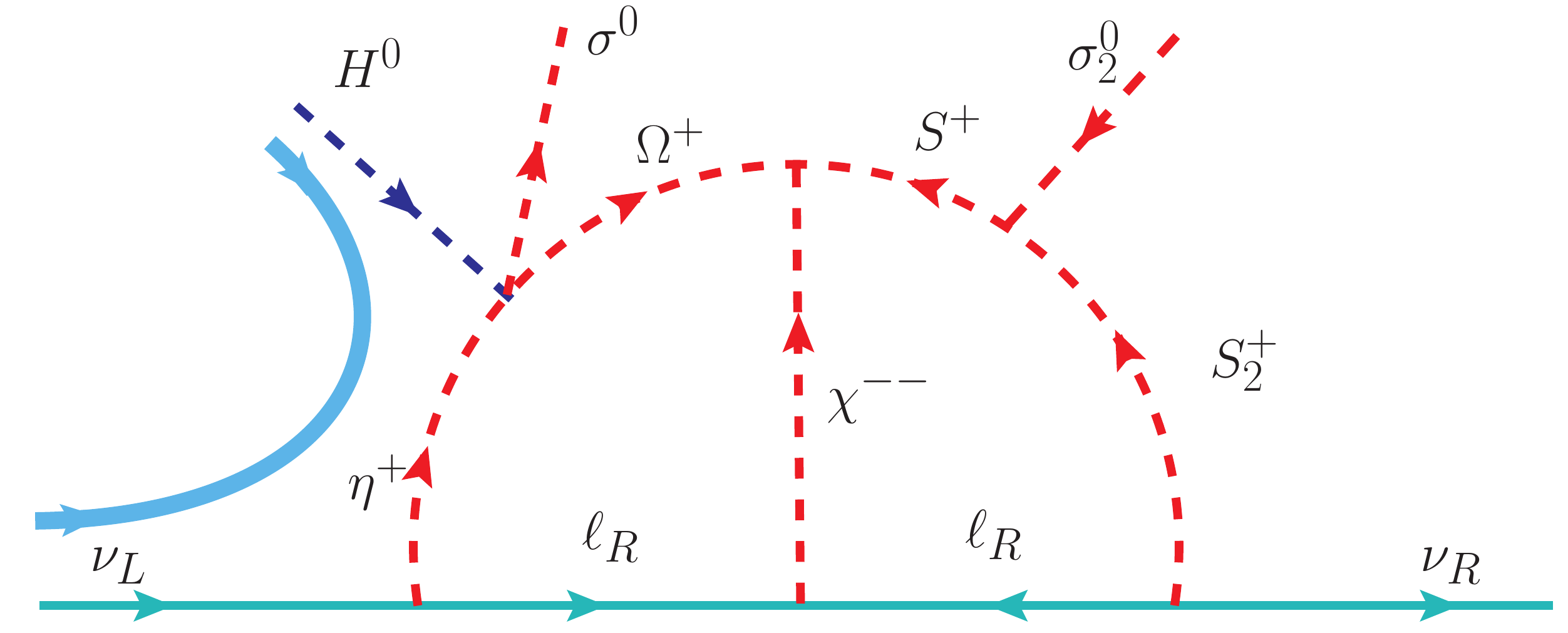}
\caption{Two-loop Feynman diagram associated with the state $\nu_{R_1}$ that carries $+5$ unit of charge under the $B-L$ for model with topology T-II-S-i. This diagram represents the UV completion of the dimension-6 operator given in Eq. \eqref{operator-6-B}. Completion of this diagram requires two extra multiplets compared to the BSM states listed in Table \ref{tab:T-II-S-i}: a neutral scalar $\sigma^{0}_2(1,0,\textcolor{blue}{9})$ and the singly charged scalar  $S^+_2(1,1,\textcolor{blue}{-4})$. }\label{dim6-3}
\end{figure} 

\noindent
Note that UV completion of this model is more economical due to the presence of a second neutral scalar $\sigma^{0}_2(1,0,\textcolor{blue}{9})$ along with the singly charged particle $S^+_2(1,1,\textcolor{blue}{-4})$. The corresponding Feynman diagram is shown in Fig. \ref{dim6-3}. To contribute to the neutrino mass, $\sigma^{0}_2$ must also acquire VEV in this scenario. We should point out that both these neutral scalars can acquire VEVs without leading to  a Goldstone in the theory. This is due to the allowed  gauge invariant quartic term of the form: $\lambda_{mix}\left(\sigma^{0}\right)^3\sigma^0_2$. As a result of the spontaneous symmetry breaking $Z^{\prime}$ will eat up the Goldstone boson and the other linear combination of the imaginary parts will receive a mass that is directly proportional to the mixed quartic coupling and which we find:  $m^2=-\lambda_{mix}\left(v_{\sigma}^2+9v_{\sigma_2}^2  \right) v_{\sigma}/\left( 2v_{\sigma_2}\right)$. However,   this mechanism  does not require $\sigma_2^0$ to acquire an explicit VEV, because the presence of the mixed quartic term as aforementioned automatically induces a VEV for it.   Hence, implementation of the operator Eq. \eqref{operator-6} requires three singly charged particle whereas, Eq. \eqref{operator-6-B} requires two: one neutral and one singly charged scalars.

\FloatBarrier
\begin{figure}[b!]
\centering
\includegraphics[scale=0.4]{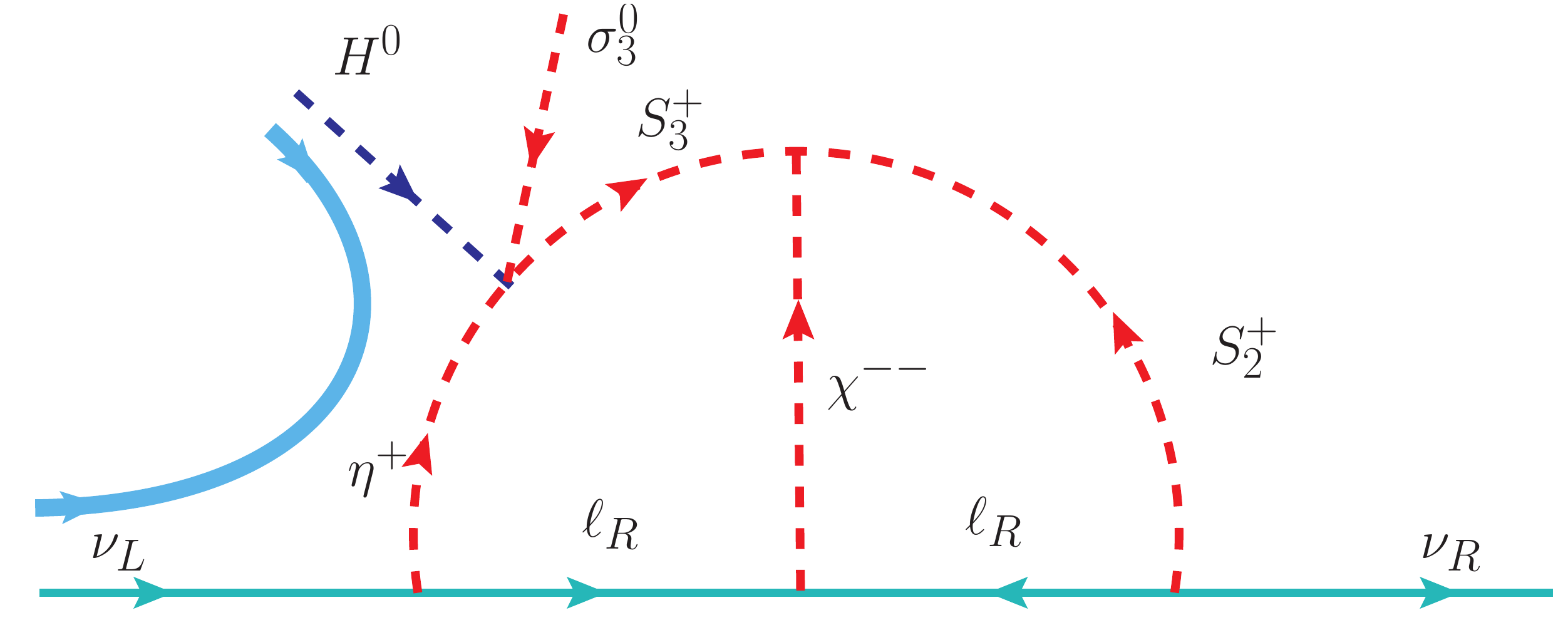}
\caption{Two-loop Feynman diagram associated with the state $\nu_{R_1}$ that carries $+5$ unit of charge under the $B-L$ for model with topology T-II-S-i. This diagram represents the UV completion of the dimension-5 operator given in Eq. \eqref{operator-5-B}. Completion of this diagram requires three extra multiplets compared to the BSM states listed in Table \ref{tab:T-II-S-i}: a neutral scalar $\sigma^{0}_3(1,0,\textcolor{blue}{6})$ and two singly charged scalars: $S^+_2(1,1,\textcolor{blue}{-4})$ and $S^+_3(1,1,\textcolor{blue}{6})$. }\label{dim6-3-C}
\end{figure} 

In this same model with topology T-II-S-i, the mass of the associated chiral states $\nu_{R_1}$ can also be lifted alternatively with a lower dimensional operator  (dimension-5)  by allowing a different neutral scalar $\sigma^0_3$ that has the following form:
\begin{align}
\mathcal{L}_5= \frac{h_{i1}}{\Lambda}\;\overline{L_L}_i \widetilde{H}\;{\nu_R}_1    \sigma^{0\ast}_3. \label{operator-5-B}
\end{align} 

\noindent
Associated with this dimension-5 operator, two-loop Feynman diagram as shown in Fig. \ref{dim6-3-C}  can be completed by introducing a second  neutral scalar $\sigma^{0}_3(1,0,\textcolor{blue}{6})$ and two singly charged scalars: $S^+_2(1,1,\textcolor{blue}{-4})$ and $S^+_3(1,1,\textcolor{blue}{6})$. So this alternative possibility is not as economical as implementing Eq. \eqref{operator-6-B}, but utilizes dimension-5 instead of dimension-6 operator.  The newly introduced neutral scalar $\sigma^0_3$ also needs to acquire a VEV (induced or explicit).  Note that, this model does not lead to any massless physical Goldstone boson either even if it were to acquire explicit VEV. This is   due to the gauge invariant cubic term in the scalar potential: $\mu_{mix}\left(\sigma^{0}\right)^2\sigma^{0\ast}_3$. Other than the Goldstone boson eaten up by the corresponding gauge boson $Z^{\prime}$, the orthogonal linear combination of the imaginary parts that becomes massive has the mass of the form: $m^2=-\mu_{mix}\left( v^2_{\sigma}-4v^2_{\sigma_3} \right)/(\sqrt{2}v_{\sigma_3})$.  If this multiplet is not given an explicit VEV, then the aforementioned cubic term automatically induces VEV for it. 

\begin{figure}[t!]
\centering
\includegraphics[scale=0.4]{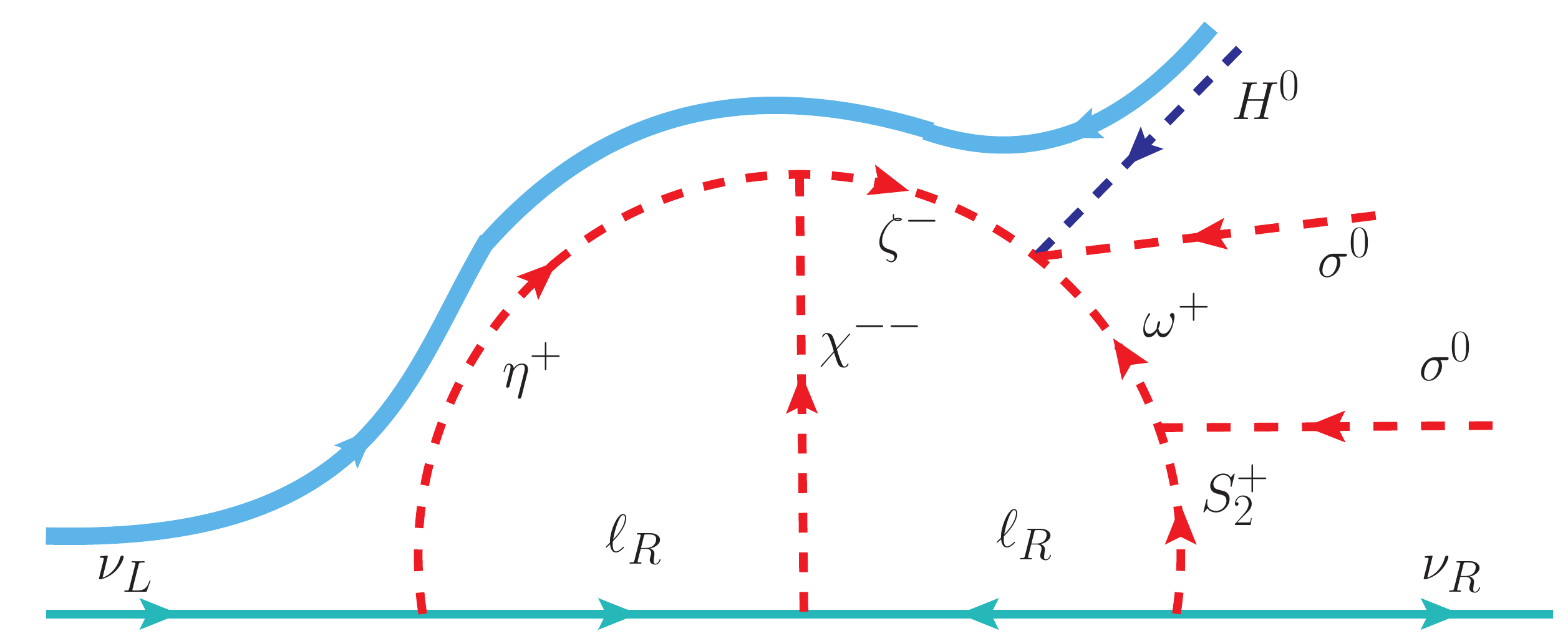}
\caption{Two-loop Feynman diagram associated with the state $\nu_{R_1}$ that carries $+5$ unit of charge under the $B-L$ for model with topology T-II-S-iii. This diagram represents the UV completion of the dimension-6 operator given in Eq. \eqref{operator-6}. Completion of this diagram requires two extra multiplets compared to the BSM states listed in Table \ref{tab:T-II-S-iii}: two singly charged  scalars:  $S^+_2(1,1,\textcolor{blue}{-4})$ and $\omega^+(1,1,\textcolor{blue}{-1})$. }\label{dim6-4}
\end{figure}

Furthermore, for two-loop model with topology T-II-S-iii, the dimension-6 operator of Eq. \eqref{operator-6} can similarly induce mass for the massless chiral states by extending the minimal model given in Table \ref{tab:T-II-S-iii} with two singly charged  scalars:  $S^+_2(1,1,\textcolor{blue}{-4})$ and $\omega^+(1,1,\textcolor{blue}{-1})$. The associated Feynman diagram is presented in Fig. \ref{dim6-4}.

\FloatBarrier

\begin{figure}[th!]
\centering
\includegraphics[scale=0.4]{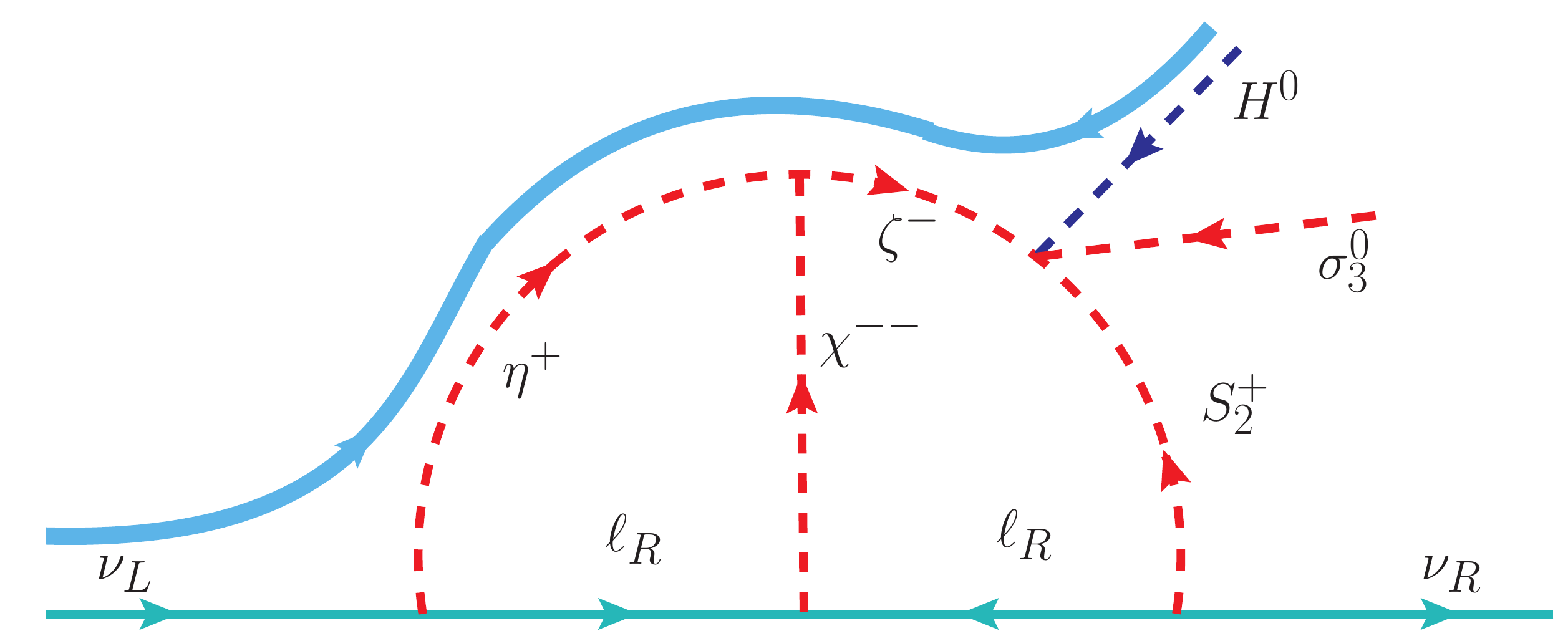}
\caption{Two-loop Feynman diagram associated with the state $\nu_{R_1}$ that carries $+5$ unit of charge under the $B-L$ for model with topology T-II-S-iii. This diagram represents the UV completion of a new  dimension-5 operator given in Eq. \eqref{operator-5-B}. Completion of this diagram requires two extra multiplets compared to the BSM states listed in Table \ref{tab:T-II-S-iii}: one singly charged  scalars:  $S^+_2(1,1,\textcolor{blue}{-4})$ and a neutral multiplet $\sigma^0_3(1,1,\textcolor{blue}{6})$. }\label{dim6-5}
\end{figure}

For the model under consideration, alternatively one can use the modified  dimension-5 operator introduced in Eq. \eqref{operator-5-B}.  
Allowance of this new operator requires the existence of a new neutral scalar $\sigma^0_3(1,1,\textcolor{blue}{6})$ along with the singly charged scalar $S^+_2(1,1,\textcolor{blue}{-4})$. The corresponding Feynman diagram is exhibited in Fig. \ref{dim6-5}. This model also does not include any massless Goldstone states due to the same reason as  explained above. 

\FloatBarrier

\section{\large Effects of higher dimensional operators: Three-loop models}\label{C} 
Similar to the two-loop models, allowed gauge invariant higher dimensional operators will induce mass to the massless chiral state associated to $\nu_{R_1}$. Here we discuss the allowed higher dimensional operators for the minimal three-loop model that has topology T-III-F-i with particle content given in Table \ref{tab:T-III-F-i}. Similar discussion can be carried out trivially for the next-to-minimal models  discussed in Sec. \ref{three}  that we do not present here. Since this three-loop model we considered is generated by the dimension-7 operator given in Eq. \eqref{operator-7},
the next to leading order operator of the following form can correct the mass for the massless states:
 \begin{align}
\mathcal{L}_8= \frac{h_{ij}}{\Lambda^4}\;\overline{L_L}_i \widetilde{H}\;{\nu_R}_j\left( \sigma^{0\ast} \right)^2 (H^{\dagger}H). \label{operator-8}
\end{align}

\FloatBarrier
\begin{figure}[t!]
\centering
\includegraphics[scale=0.4]{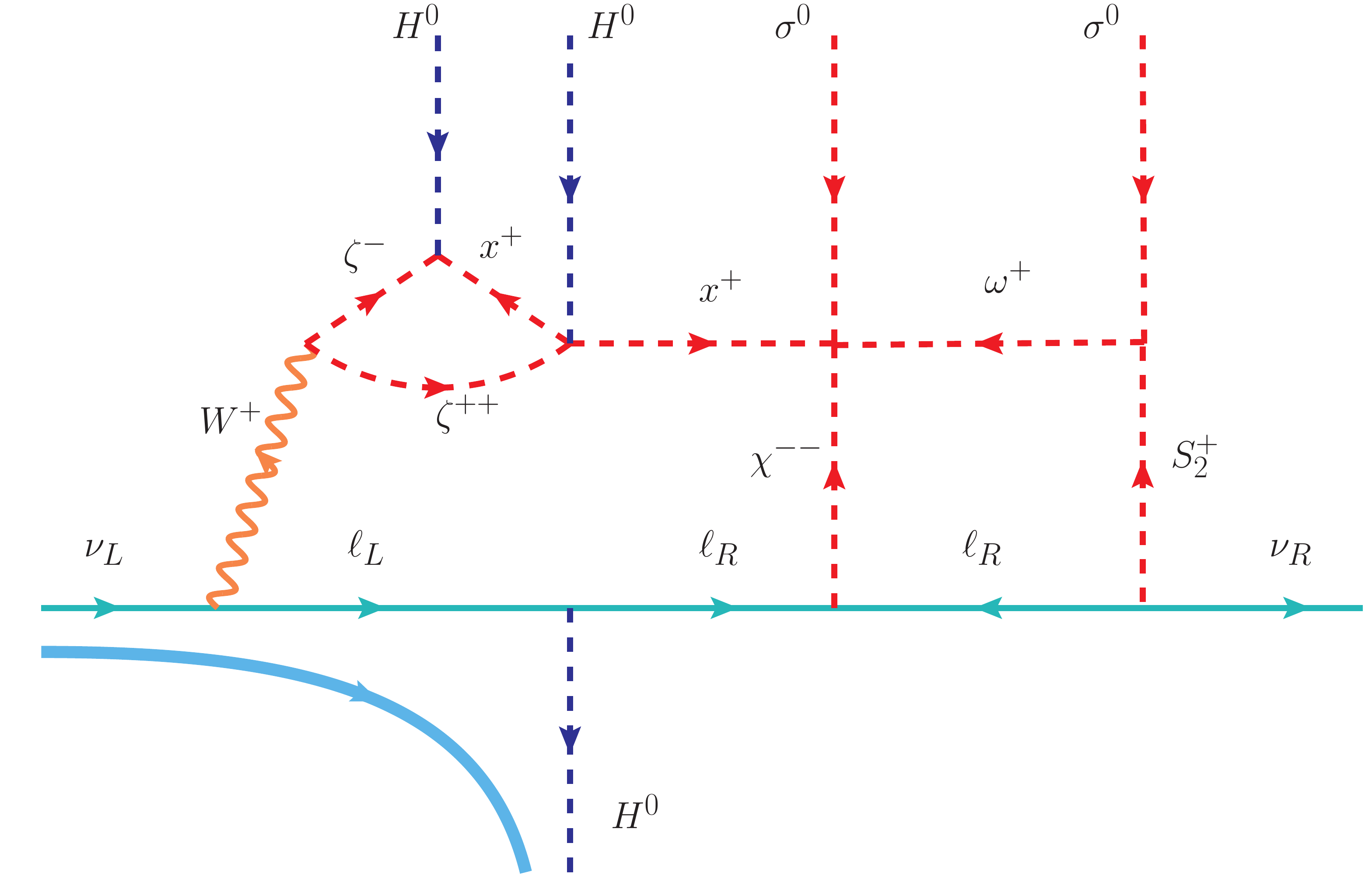}
\caption{Three-loop Feynman diagram associated with the state $\nu_{R_1}$ that carries $+5$ unit of charge under the $B-L$ for model with topology T-III-F-i. This diagram represents the UV completion of the  dimension-8 operator allowed within the minimal model given in Eq. \eqref{operator-8}. Completion of this diagram requires two extra singly charged scalars compared to the BSM states listed in Table \ref{tab:T-III-F-i}:  $S^+_2(1,1,\textcolor{blue}{-4})$ and $\omega^+(1,1,\textcolor{blue}{-1})$.  }\label{dim8}
\end{figure} 

\noindent
This dimension-8 operator will lead to the Feynman diagram shown in Fig. \ref{dim8}. The UV completion of this operator requires the introduction of  two singly charged multiplets: $S^+_2(1,1,\textcolor{blue}{-4})$ and $\omega^+(1,1,\textcolor{blue}{-1})$ in addition to the already existing BSM scalar multiplets listed in Table \ref{tab:T-III-F-i}.  

\FloatBarrier
\begin{figure}[t!]
\centering
\includegraphics[scale=0.4]{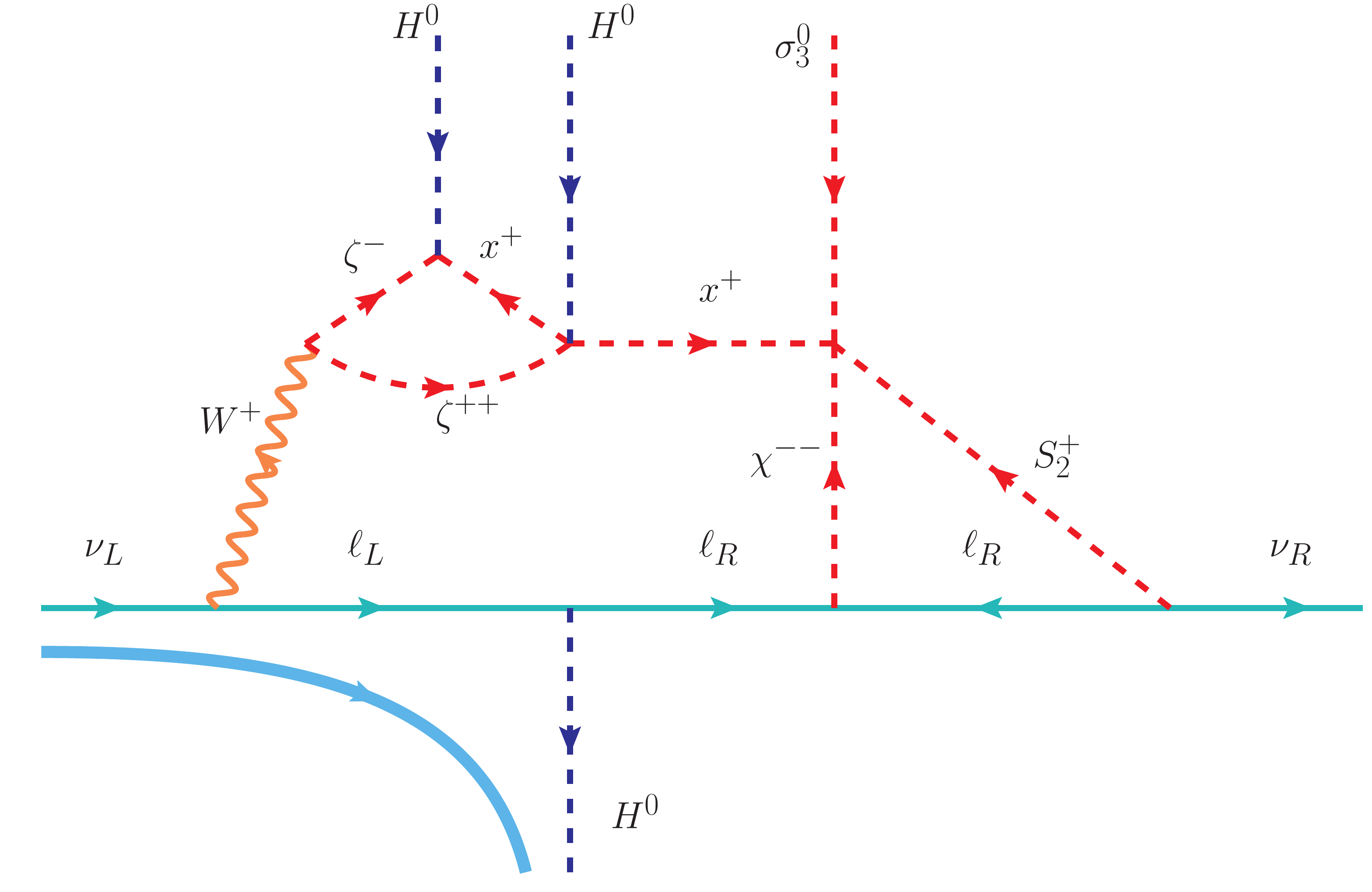}
\caption{Three-loop Feynman diagram associated with the state $\nu_{R_1}$ that carries $+5$ unit of charge under the $B-L$ for model with topology T-III-F-i. This diagram represents the UV completion of a new  dimension-7 operator given in Eq. \eqref{operator-7-B}. Completion of this diagram requires two extra multiplets compared to the BSM states listed in Table \ref{tab:T-III-F-i}:  a neutral member $\sigma^0_3(1,1,\textcolor{blue}{6})$ and a charged member $S^+_2(1,1,\textcolor{blue}{-4})$. }\label{dim7}
\end{figure} 

An alternative way to induce this mass within dimension-7 is to introduce a second SM singlet $\sigma^{0}_3$ that allows the following new operator originally not present in the minimal set-up:  
 \begin{align}
\mathcal{L}_7= \frac{h_{ij}}{\Lambda^3}\;\overline{L_L}_i \widetilde{H}\;{\nu_R}_j \sigma^{0}_3 (H^{\dagger}H). \label{operator-7-B}
\end{align}  

\noindent 
The UV completion of this model requires two extra multiplets as well which are: a singly charged particle $S^+_2(1,1,\textcolor{blue}{-4})$ and a neutral scalar $\sigma^0_3(1,1,\textcolor{blue}{6})$. As already pointed out above, acquiring a VEV by this newly introduced neutral member does not lead to any massless Goldstone boson in the theory. The corresponding Feynman diagram is presented in Fig. \ref{dim7}.

\section{\large Non-minimality of T-III-S topology}\label{non-min}
In Sec. \ref{three}, we argued that three-loop models constructed  from T-III-F-x topology (Fig. \ref{three-loop}) are more economical compared to T-III-S-x topology (Fig. \ref{three-loop-B}) introduced in Sec. \ref{three}. In this appendix, we explicitly build minimal models within  topology of type T-III-S-x for a comparison.

\begin{figure}[t!]
\centering
\includegraphics[scale=0.43]{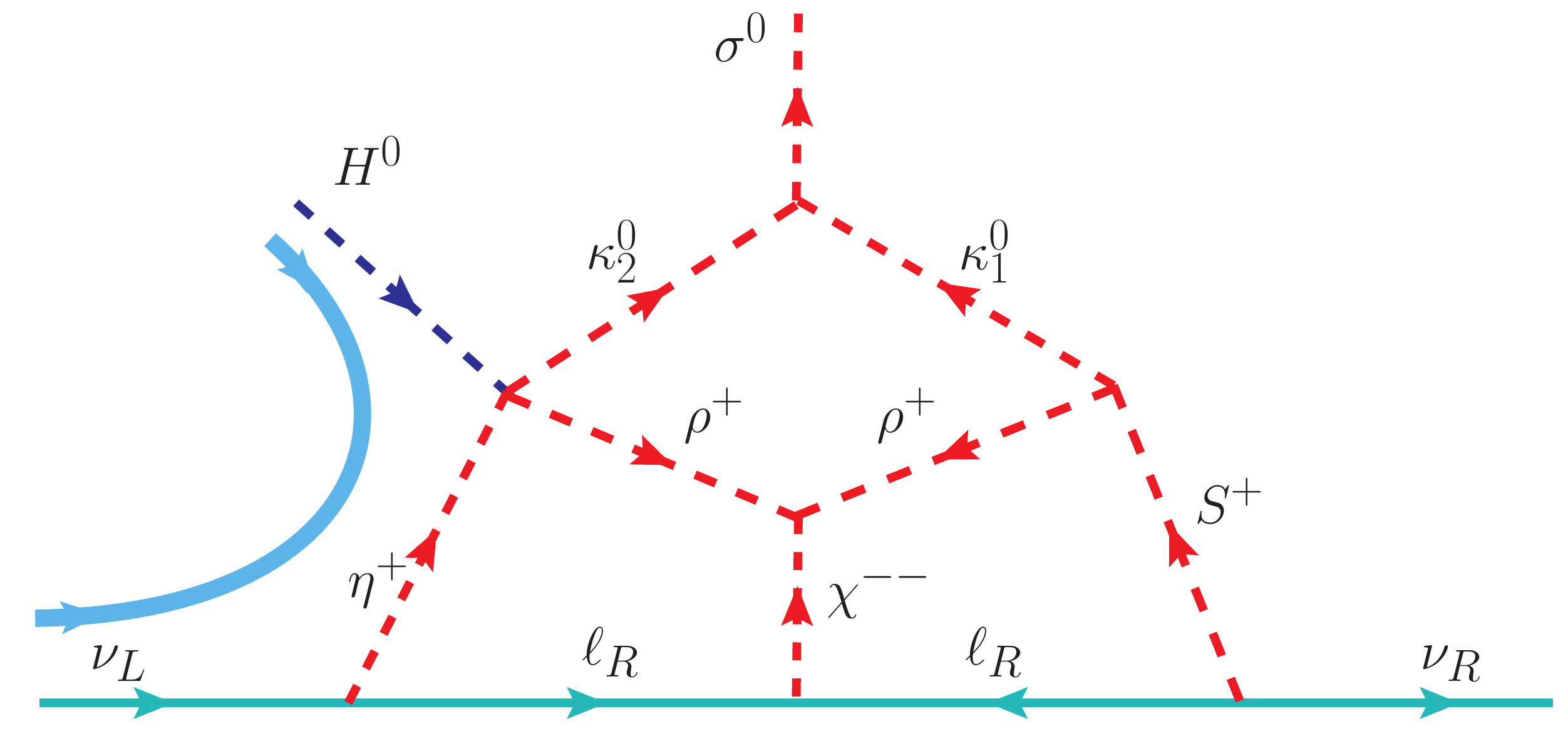}
\caption{Three-loop Dirac neutrino mass for the particle content shown in Table \ref{tab:T-III-S-i}. }\label{fig:T-III-S-i}
\end{figure}

\begin{figure}[th!]
\centering
\includegraphics[scale=0.43]{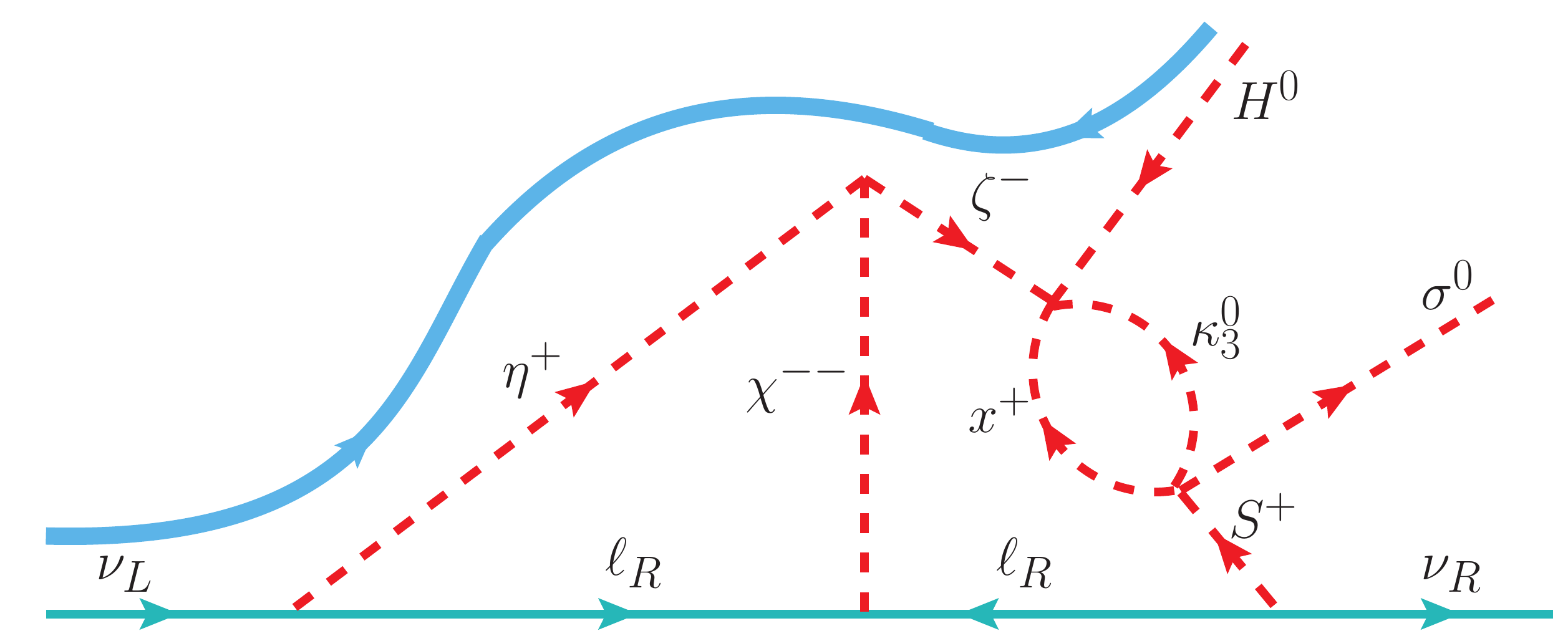}
\caption{Three-loop Dirac neutrino mass for the particle content shown in Table \ref{tab:T-III-S-iii}. }\label{fig:T-III-S-iii}
\end{figure}

First we remind the readers that  diagrams of this type (T-III-S-x shown in Fig. \ref{three-loop-B}) do not   use any SM gauge bosons running in the loop, hence  have the same basic features as that of the two-loop topologies (T-II-S-x) given in Fig. \ref{two-loop}.  Hence according to our definition, these three-loop topologies should be labeled as: T-III-S-i, T-III-S-ii  and T-III-S-iii depending on how the iso-spin doublet flows within a given diagram following Fig. \ref{two-loop}. As already pointed out, T-II-S-ii topology   is not interesting as its  implementation automatically induces  one-loop contribution, hence cannot be a true two-loop model (same argument goes for T-III-S-ii). So below, we consider T-III-S-i and T-III-S-iii topologies. 
Construction of the minimal model with topology T-III-S-i (T-III-S-iii) is  presented in Table \ref{tab:T-III-S-i} (\ref{tab:T-III-S-iii})  and the corresponding Feynman diagram in Fig. \ref{fig:T-III-S-i} (\ref{fig:T-III-S-iii}). Both these models demand no less than seven BSM scalars, no further economical model is found within these topologies. From our discussion, it is clear that topologies of type T-III-S cannot be the candidate for minimal three-loop models, since at least  seven BSM scalar multiplets are required. On the other hand, topology of the type T-III-F  requires only five BSM multiplets as discussed in Sec. \ref{three}.

\FloatBarrier
\begin{table}[t!]
\centering
\footnotesize
\resizebox{0.5\textwidth}{!}{
\begin{tabular}{|c|c|}
\hline
Topology &$SU(2)_L\times U(1)_Y\times U(1)_{B-L}$  \\ \hline\hline
T-III-S-i&
\pbox{10cm}{
\vspace{2pt}
$\sigma^0(1,0,\textcolor{blue}{3})$\\
$S^+(1,1,\textcolor{blue}{5})$\\
$\chi^{++}(1,2,\textcolor{blue}{2})$\\
$\eta(2,\frac{1}{2},\textcolor{blue}{0})$\\
$\rho^+(1,1,\textcolor{blue}{1})$\\
$\kappa^0_{1}(1,0,\textcolor{blue}{4})$\\
$\kappa^0_{2}(1,0,\textcolor{blue}{-1})$
\vspace{2pt}}
\\ \hline
\end{tabular}
}
\caption{ 
Quantum numbers of the BSM scalars.
}\label{tab:T-III-S-i}
\end{table}

\FloatBarrier
\begin{table}[th!]
\centering
\footnotesize
\resizebox{0.5\textwidth}{!}{
\begin{tabular}{|c|c|}
\hline
Topology &$SU(2)_L\times U(1)_Y\times U(1)_{B-L}$  \\ \hline\hline
T-III-S-iii&
\pbox{10cm}{
\vspace{2pt}
$\sigma^0(1,0,\textcolor{blue}{3})$\\
$S^+(1,1,\textcolor{blue}{5})$\\
$\chi^{++}(1,2,\textcolor{blue}{2})$\\
$\eta(2,\frac{1}{2},\textcolor{blue}{0})$\\
$\zeta(2,-\frac{3}{2},\textcolor{blue}{-2})$\\
$x^+(1,1,\textcolor{blue}{0})$\\
$\kappa^0_{3}(1,0,\textcolor{blue}{2})$
\vspace{2pt}}
\\ \hline
\end{tabular}
}
\caption{ 
Quantum numbers of the BSM scalars.
}\label{tab:T-III-S-iii}
\end{table}
\FloatBarrier

\end{appendices}

\FloatBarrier

\end{document}